\documentclass[12pt]{article}

\title{Uncovering simultaneous breakthroughs with a robust measure of disruptiveness}
\usepackage{amsmath,amssymb}
\usepackage{authblk}
\usepackage[margin=1in]{geometry}
\usepackage{url}
\usepackage{subfigure}
\usepackage{amsmath}
\usepackage{listings}
\usepackage{nameref}
\usepackage{hyperref}
\hypersetup{hidelinks}
\author[1]{Munjung Kim}
\author[2]{Sadamori Kojaku}
\author[1]{Yong-Yeol Ahn}
\affil[1]{Center for Complex Networks and Systems Research, Luddy School of Informatics, Computing, and Engineering,
Indiana University, Bloomington}
\affil[2]{School of Systems Science and Industrial
Engineering, Binghamton University, State University of New York}
\affil[1]{\textit{yyahn@iu.edu}}

\usepackage[dvipsnames]{xcolor}
\usepackage{soul}
\usepackage{geometry}
\usepackage{booktabs}
\usepackage{caption}
\usepackage{tabularx}
    \newcolumntype{L}{>{\raggedright\arraybackslash}X}

\usepackage{rotating}

\usepackage{graphicx}
\usepackage{inputenc}
\usepackage{setspace} 
\usepackage[sorting=none, backend=biber,isbn=false,alldates=year,
eprint=false,url=false, style=nature]{biblatex}
\AtEveryBibitem{%
  \clearfield{issue}%
}
\usepackage{amsmath}

\newcommand{\Dnok}{D_{\text{no}k}}

\begin{document}

\maketitle

\begin{abstract}

Progress in science and technology is punctuated by disruptive innovation and breakthroughs. Researchers have characterized these disruptions to explore the factors that spark such innovations and to assess their long-term trends. However, although understanding disruptive breakthroughs and their drivers hinges upon accurately quantifying disruptiveness, the core metric used in previous studies---the disruption index---remains insufficiently understood and tested. Here, after demonstrating the critical shortcomings of the disruption index, including its conflicting evaluations for \emph{simultaneous discoveries}, we propose a new, continuous measure of disruptiveness based on a neural embedding framework that addresses these limitations. Our measure not only better distinguishes disruptive works, such as Nobel Prize-winning papers, from others, but also reveals simultaneous disruptions by allowing us to identify the ``twins'' that have the most similar future context. By offering a more robust and precise lens for identifying disruptive innovations and simultaneous discoveries, our study provides a foundation for deepening insights into the mechanisms driving scientific breakthroughs while establishing a more equitable basis for evaluating transformative contributions.

\end{abstract}
\doublespacing

\newpage
\section{Introduction}

A perennial dichotomy in explaining the progress of science has been between ``developing'' contributions---small, marginal, continuous, and ``normal'' progresses---versus ``disruptive'' ones---big, discontinuous leaps~\cite{kuhn2012structure, schumpeter2013capitalism, arthur2009nature,wu2019large, funk2017dynamic}. From this perspective, most works are \emph{marginal}, realizing logically expected next steps and extending the existing streams of works, while there are exceptions that \emph{disrupt} existing streams and make prior works obsolete by imagining a new theory, concept, or way of thinking.

\begin{figure}
    \centering
    \includegraphics[width=0.9\textwidth]{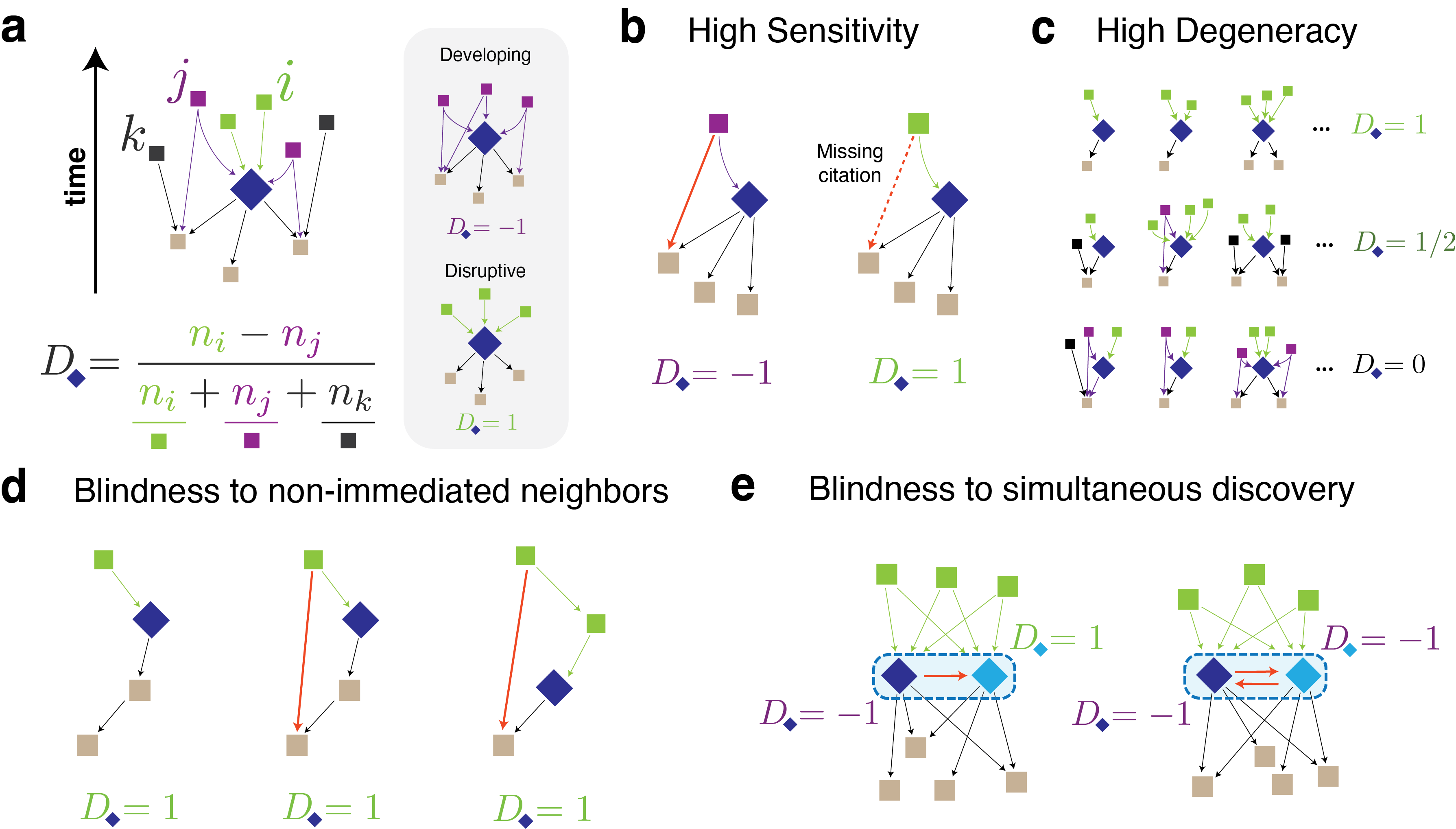}
    \caption{ \textbf{The disruption index has critical limitations due to its discreteness and locality.} \textbf{(a)} The disruption index quantifies the degree to which the descendent works rely solely on a focal work and are free from antecedent works. The disruption index of a focal paper reaches its minimum value $D=-1$ when all future works citing the focal work also cite the prior works referenced by the focal work. On the other hand, the focal work is maximally disruptive ($D=1$) when all future works cite the focal work while not citing any past works referenced by the focal work.  
   \textbf{(b)} 
   The disruption index is extremely sensitive; even a single missing citation can cause it to shift dramatically from -1 to 1.
    \textbf{(c)} $D$ exhibits high degeneracy, having the same values for the different citation topology structures. 
    \textbf{(d)} The index only captures the local structure formed by directly connected papers, neglecting any structure beyond the immediate vicinity.   \textbf{(e)} In cases where two papers jointly or simultaneously create disruption and receive equal recognition from descendants, even a single citation can turn them minimally disruptive from maximally disruptive. }
    \label{fig:problems}
\end{figure}

Although many tried to capture this fundamental tension~\cite{uzzi2013atypical,kim2022quantifying,milojevic2015quantifying}, the most successful operationalization is, arguably, the \emph{disruption index}~\cite{funk2017dynamic,wu2019large}. The disruption index builds on the idea that disruptive works should \emph{replace} the established practices, making previous works obsolete. 
As a result, a disruptive work is assumed to create a bow-tie structure---a bottleneck---in the citation network, where it \emph{severs} the stream of marginal innovations to create a new stream. The disruption index of a paper aims to quantify the clarity of this bow-tie structure by contrasting the number of subsequent works that build only on the focal work against the number of other subsequent works that build on both focal work and its referenced prior works (see Figure~\ref{fig:problems}a). The disruption index has been widely adopted, leading to discoveries of factors that may induce the creation of more disruptive innovations including team size~\cite{wu2019large}, field size~\cite{chu2021slowed}, atypical knowledge combinations~\cite{lin2022new}, technological niches~\cite{qu2024outliers}, remote or fresh collaboration~\cite{lin2023remote,zeng2021fresh},  conceptual structure~\cite{kedrick2024conceptual}, and increasing reliance on specialized knowledge~\cite{park2023papers}.

The disruption index is, however, a \emph{discrete} measure that relies on the topology of the local citation network, hampered by debilitating limitations (see Figure~\ref{fig:problems}). First, the disruption index can be extremely sensitive to small---even a single citation---changes in the topology of the local citation network (see Figure~\ref{fig:problems}b). Second, the index has a low resolution because it \emph{counts} the papers that are classified into just three categories without taking into account any other information. For instance, even if the descendent paper relies on the focal paper citing more antecedent papers, the index does not change. Therefore, the index is more likely to have specific ratio values (e.g., 0, 1, and $\frac{1}{2}$), leading to high degeneracy (see Figure~\ref{fig:problems}c).  Third, because the index is calculated by examining only one step of the citation, it cannot capture any information that is not encoded in the immediate vicinity of the focal paper (see Figure~\ref{fig:problems}d). 

Finally, the disruption index can completely obscure \emph{simultaneous disruption} events (see Figure~\ref{fig:problems}e). Simultaneous disruption occurs when scientists independently make similar, or even the same, discoveries without knowledge of each other's work, or when the same author(s) publish multiple papers that collectively represent a single breakthrough~\cite{merton1961singletons, bikard2020idea}.  Since the ``paper'' is the unit of analysis, these ``multiples'' that represent the same discovery are treated as separate entities when the disruption index is applied. This approach overlooks simultaneous disruptions and can even render them \emph{invisible}. Specifically, when one paper contributing to this simultaneous disruption cites another, the disruption index mistakenly interprets \emph{all} subsequent studies that cite both papers as descendants relying on the references of the collective disruption papers. This misclassification shifts papers that should be counted as $n_i$ to $n_j$, drastically reducing the estimated disruptiveness. In the most extreme case where two publications represent a perfectly disruptive innovation where none of the descendants cite any of the antecedents, with the addition of a \emph{single} citation, the disruption index can change from 1 (maximally disruptive) to -1 (minimally disruptive). 
In our analysis, we find numerous instances of simultaneous disruptions that were obscured by this issue.

\begin{figure}
    \centering
\includegraphics[width=0.82\textwidth]{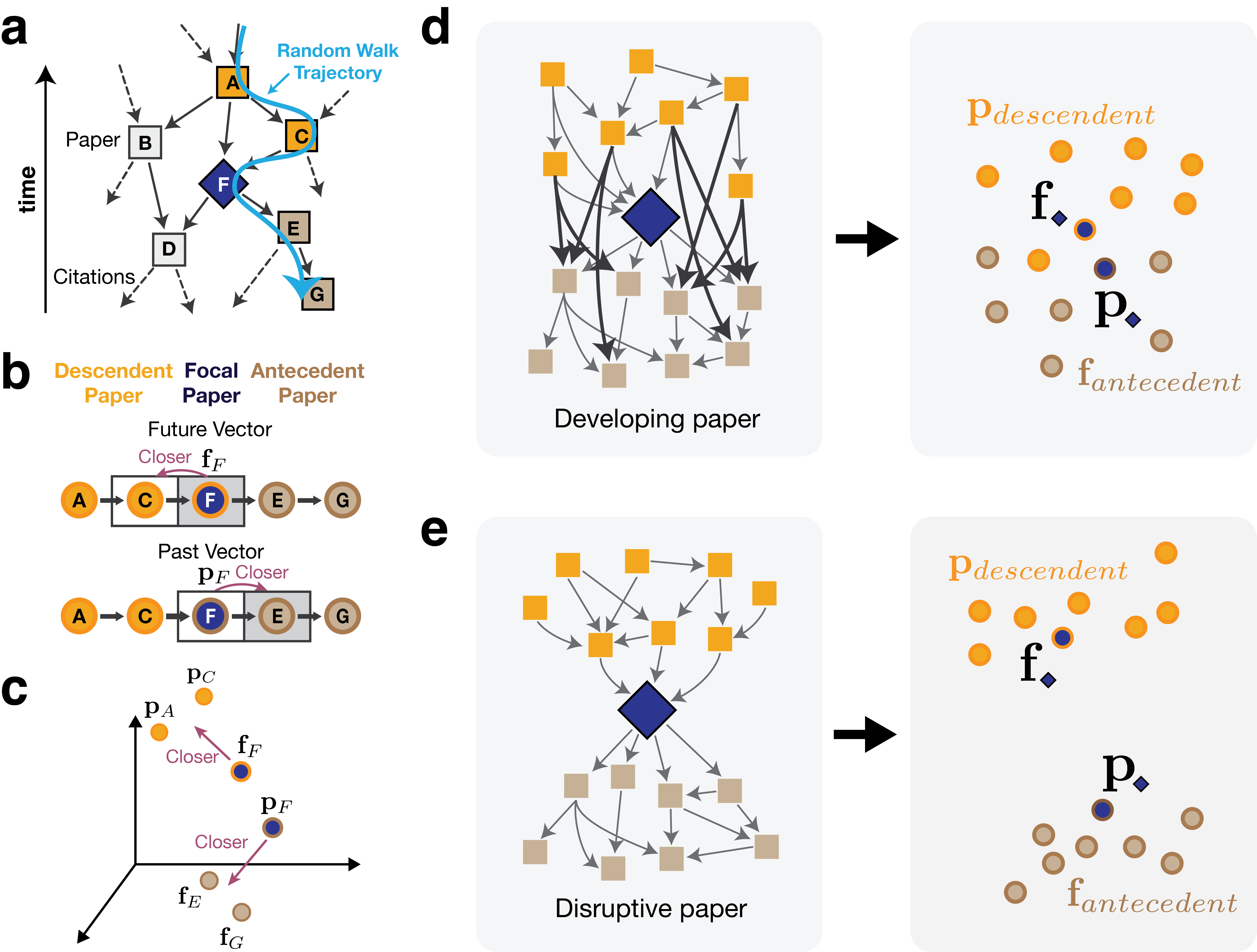}
    \caption{\textbf{Directional graph embedding captures disruptiveness.}  Unlike the disruption index, our embedding approach leverages the entire network structure to estimate the disruptiveness of each paper. This approach separately represents the citing and cited features of papers (see Methods for a detailed explanation of the algorithm). \textbf{(a)} First, we generate random walks (blue arrow) on the citation network. \textbf{(b)} Our model aims to learn two vectors (``future'' and ``past'') for each paper that can be used to accurately predict `what comes before' (future vector) and `what comes next' (past vector) in the random walk trajectories. \textbf{(c)} As a result, future vector $\mathbf{f}$ approaches descendent papers vectors while past vector $\mathbf{p}$ approaches the antecedent papers vectors. \textbf{(d)} For the developing paper,  the distance between the vectors representing antecedent works and descendent works are close in the embedding space because of the large reliance of descendent works on antecedent works. This makes the distance between future vector $\mathbf{f}$ and past vector $\mathbf{p}$ becomes closer. \textbf{(e)}  For the disruptive paper, on the other hand, the distance between future vector and past vector becomes greater, as the fewer connections between antecedent papers and descendent papers make their representation vector far away in the latent space.}
    \label{fig:disruptiveschematics}
\end{figure}

Simultaneous discoveries may not be outliers but commonplaces~\cite{merton1961singletons,simonton1979multiple}. Newton and Leibniz independently formulated calculus; Wallace and Darwin simultaneously devised the theory of evolution, and multiple researchers independently discovered the periodic table and energy conservation. The Nobel Prize, the most prestigious award in science, often honors multiple individuals, sometimes even sparking controversy of crediting limited people among many contributors~\cite{merali_physicists_2010}. Beyond a few examples involving prominent figures in scientific research, many scholars have proposed the prevalence of simultaneous discoveries across general science, arguing that simultaneous discoveries are a common, yet largely overlooked pattern of scientific discoveries, partly due to the lack of the methods that can discover them systematically~\cite{simonton1979multiple,merton1961singletons}. Therefore, the inability to recognize simultaneous discoveries can hinder our understanding of scientific progress.

The limitations of the disruption index call for a continuous and more robust measure of disruptiveness. Here, we propose a continuous measure that addresses these limitations by combining the conceptual foundation of the disruption index with a method from neural language models and graph embeddings. Our approach builds on the premise that, just like neural language models that represent entities (words) as vectors in a high-dimensional vector space, we can imagine each paper as a vector that captures the particular subject of the study. Then, embracing the basic idea of the disruption index---where a disruptive paper diminishes the relevance of earlier works for future research---we develop a method to train two distinct vectors for each paper in the same embedding space: one representing the context of ``ancestors,'' and the other representing the context of ``descendants.'' Disruptive works trigger a contextual shift that creates a new stream of works. Thus, we postulate that the distance between these two representations reflects the extent to which ``descendants'' departed from its ``ancestors.'' We also provide a mathematical formulation demonstrating how this distance effectively captures the extent to which descendants rely on their ancestors---the phenomenon that the disruption index was originally designed to measure (see Methods).

Using a dataset of over 55 million general scientific papers from Web of Science dataset and physics-related papers from American Physics Society dataset, we demonstrate that our measure addresses the shortcomings of the disruption index. Not only does it more accurately identify disruptive works, such as Nobel Prize-winning papers, but it can also discover simultaneous disruptions that the disruption index fails to recognize.

\section{Results}
\label{sec:result}

The key idea behind our approach is that we can imagine, for each paper, \emph{two} distinct vectors: the \emph{past} and \emph{future}. The past vector $\mathbf{p}_i$ of paper $i$, points to the antecedent papers---its references and their close relatives, whereas the future vector $\mathbf{f}_i$ points to the future relatives of the focal paper. If the embedding model is trained such that the proximity between the vectors indicates higher connections between their papers, and if disruptive papers indeed tend to eclipse the future knowledge from the past, making future knowledge less rely on the past, we expect that a paper’s past and future vectors diverge as the paper’s disruptiveness increases. Thus, by quantifying the distance between these two vectors---representing the past and future context of each paper---we can estimate their disruptiveness. Because they are defined in the continuous embedding space from the entire citation network structure, rather than just the 1-hop neighborhood of the focal paper, we can also expect that this measure is more robust, incorporating higher-order relationships, reducing degeneracy, and remaining stable against individual citation link changes.

To accomplish this, let us imagine an embedding framework where two representation vectors are learned from the past and future contexts respectively, based on the citation relationship. One way to achieve this is to employ an objective function that aligns the \emph{past} vector of the focal paper with the \emph{future} vectors of its antecedent papers, and the \emph{future} vector of the focal paper with the \emph{past} vectors of its descendant papers. Then, we can create training examples by generating direction-aware random walks from the citation network (Figure~\ref{fig:disruptiveschematics} a). Mathematically, by employing a commonly-used softmax-based formulation, we can first assume that, if we obtain good representations for every paper, the conditional probability of seeing paper $j$ after seeing paper $i$ (within a short window) on a random walk trajectory that follows the citations to the past can be written as 
\begin{equation}
    \text{Pr}(j|i) = \frac{\exp{(\mathbf{f_j} \cdot \mathbf{p_i})}}{\sum_k \exp{\mathbf{f_k} \cdot\mathbf{p_i}} },
\end{equation}
where $k$ is all papers in the dataset. We then use the skip-gram objective function

\begin{equation}\label{eqn:objective_randomwalk}
\mathcal{J} = \sum_{s \in V} \sum_{r=1}^{R} \sum_{t=1}^{T - c} \sum_{w=1}^{c} \log \text{Pr}(v_{t+w}^{(r, s)} | v_t^{(r, s)}),
\end{equation}

where $V$ represents the set of all papers, and $s \in V$ denotes a paper from which a random walk is initiated. For each paper $s$, $R$ random walks are performed, where $r$ indexes the random walks, and $t$ is the position within each random walk of length $T$. The context window size is denoted by $c$, and $w$ represents the citation step within the window. The term $v_t^{(r, s)}$ refers to the node at position $t$ in the $r$-th random walk starting from paper $s$, while $v_{t+w}^{(r, s)}$ is the antecedent paper of $v_t^{(r, s)}$ within $w$ citation steps. 

When $R$ and $T$ are big enough, Equation (\ref{eqn:objective_randomwalk}) can be approximated as 
\begin{equation}\label{eqn:objective_intro}
\mathcal{J} \approx    \sum_{u \in V} \sum_{v\in A_c(u)} \kappa_u^{\text{in}}\log \text{Pr}(v | u),
\end{equation}
where $A_c(u)$ is the set of antecedent papers that are cited by the paper $u$ within $c$ citation steps. Here, the term $\log \text{Pr}(v|u)$ is weighted by $\kappa_u^{\text{in}}$, which is the in-degree (number of citations) of the paper $u$, as the density of a random walker can be approximated as proportional to the in-degree of a node of a network~\cite{masuda2017random, fortunato2007random}. We learn the vectors that maximize this likelihood function based on many random work trajectories sampled from the citation network. In this way, one type of vector, $\mathbf{f}$, learns which papers are likely to be descendent works connected within $c$ citation steps, while the other type of vector, $\mathbf{p}$, learns which papers are likely to be prior works within $c$ citation steps (Figure~\ref{fig:disruptiveschematics}b and c). This approach is equivalent to the node2vec model~\cite{grover2016node2vec} (or the word2vec model) applied to citation trajectories treated as `sentences'~\cite{mikolov2013distributed}, but with the window constrained to a single direction.

When our algorithm is applied to a citation network, the cosine distance between the future vector and the past vector of a paper reflects reliance (or the lack thereof) of its descendent works on its antecedent works (Figure~\ref{fig:disruptiveschematics}d; see Methods for the mathematical description of this argument.) Thus, we introduce Embedding Disruption Measure (EDM) index for paper $i$ as the cosine distance between the past vector (representing references, $\mathbf{p}_i$) and the future vector (representing citations, $\mathbf{f}_i$) of paper $i$:

\begin{equation}
\Delta_i =1-\frac{\mathbf{f}_i\cdot \mathbf{p}_i}{| \mathbf{f}_i| |\mathbf{p}_i|}.
\end{equation}

We choose to use cosine distance instead of Euclidean distance because the vector norms reflect the frequency of the papers appearing in the random walk sequences, or the `attractiveness' of the paper---the information we specifically want to exclude from our measure~\cite{murray2020unsupervised,schakel2015measuring}.

\subsection{Degeneracy and Locality}

First, let us show the distribution of the measures across all papers in Web of Science and American Physical Society datasets (see Methods for the details of the dataset).
Given that our measure (denoted as $\Delta$) is continuously defined and free from degeneracy issues, we expect that our measure exhibits a smoother distribution with a greater resolution. This expectation is clearly demonstrated in Figure~\ref{fig:distribution_disruption}a. While the $D$ tends to bunch at zero with high degeneracy, $\Delta$ exhibits a much greater resolution and a smooth distribution. The $D$'s tendency to produce values very close to zero is due to $n_k$ term, which tends to be significantly larger than other terms in the index, as it represents the total number of papers citing at least one of the references of the focal paper~\cite{bornmann2020disruption}. Since $n_k$ is generally large and located in the denominator, it dominates the index and makes the scores close to zero. We further examined the $D_{nok}$ index, where the $n_k$ term is omitted, and the $D_5$ index, which only considers citations occurring 5 years after the publication of the focal paper. Our analysis confirmed that both variants still exhibit high degeneracy (see Supplementary Section 1 for the details).

A higher resolution of $\Delta$ is also linked to its ability to capture a broader scope of information beyond a single citation. Even for the papers that share the same citation pattern within a 1-hop citation network with the same $D$ value, their $\Delta$ values can differ, as their 2- or 3-hop citation networks exhibit different patterns. We additionally explored whether $\Delta$ incorporates higher-order citation patterns by comparing $D$ computed over two citation steps with $\Delta$ in Supplementary Section 2.

\subsection{Identification of Disruptive Papers}
\label{sec:result_identification}
We compare the performance of EDM and the disruption index in identifying disruptive papers. We focused on two distinct sets of papers, each with unique characteristics: Nobel Prize-winning papers, and ``Milestone" papers selected by American Physical Society (APS).

\subsubsection{Nobel Prize and Milestone Paper}

Let us first examine the Nobel Prize-winning papers and Milestone papers. The Nobel Prize, the most prestigious international science award, is often associated with groundbreaking breakthroughs in various fields~\cite{NSF_nobel}. Similarly, milestone papers, selected by APS editors for their significant discoveries or for pioneering new fields, are linked to transformative achievements in physics and and are often recognized as a proxy for disruptive work~\cite{bornmann2021convergent,bittmann2021applied}. In total, our analysis includes 302 Nobel Prize-winning papers identified within the WOS dataset and 278 milestone papers from the APS dataset  (see Methods for details).

As we can see in Figure~\ref{fig:distribution_disruption}b, although both $\Delta$ and $D$ tend to assess these landmark papers as highly disruptive, the disruption index produces a bimodal distribution, indicating that some papers are identified as \emph{not at all} disruptive. However, the raw index does not tell us a full story because the index can be confounded by other characteristics such as the number of references and citations. Indeed, previous studies showed that the disruption index can be affected by the number of references and exponentially increasing publications and citations ~\cite{ruan2021rethinking,petersen2023disruption,bentley2023disruption}. This means any disparity found in these analyses can be due to the number of citations and references rather than reflecting the disruptive or developmental qualities of papers. To address this issue, we compute the same indices in a randomized citation network, where we maintain the number of citations and references for each paper, as well as citation year gaps (see Methods for a detailed explanation of the construction of the randomized citation network). In doing so, we isolate what the indices capture beyond the number of citations and references. 

Figure~\ref{fig:distribution_disruption}b shows the results for the randomized network. Notably, the Nobel Prize papers in the randomized network have $D$ scores \emph{more} heavily concentrated in the top 10\% compared to the original network, indicating that their high $D$ scores can be explained largely by the number of citation and reference counts. Indeed, when we compare individual papers, as shown in Figure~\ref{fig:distribution_disruption}c, the percentile ranking of $D$ scores often remain nearly the same across the randomized and original networks, further suggesting that the null model explains their high $D$ scores. (In some cases, $D$ scores drastically shift to lower values---an observation we will explore more fully in the Results.) On the other hand, $\Delta$ values in the randomized network are much lower than those in the original network and exhibit varied changes with less pronounced patterns across the two networks, indicating that  $\Delta$ captured the disruptive quality of these papers beyond what can be attributed to citation and reference counts alone. The same pattern is observed for the APS’s milestone papers.

We further conduct multivariate logistic regression to assess the extent to which the $\Delta$, $D$, and citation count are related to the likelihood of a paper being a Nobel Prize-winning or milestone paper. To address the sparsity of these papers, we used Firth's logistic regression. For comparability of effect sizes, each percentile was scaled by dividing by 10, aligning it with the citation logarithm scale, which has a maximum value of 12.42 in WOS papers and 9.28 in APS papers.

The regression analysis reveals that $\Delta$ exhibits a much stronger association with milestone papers and Nobel prize-winning papers. As shown in Figure~\ref{fig:distribution_disruption}d, for milestone papers, the odds ratios of the $\Delta$ percentile and the logarithm value of citations are 1.23 ($p<0.001$; 95\% CI: 1.14--1.34) and 6.14 ($p<0.001$; 95\% CI: 5.51--6.87), respectively. This implies that when the percentile of $\Delta$ increases by 10\%, the odds of a paper being a milestone increase by 1.23. Similarly, when the number of citations increases by about two times, the odds of a paper being a milestone increase by 6.14. Conversely, the 95\% confidence interval for the odds ratio of the D percentile includes 1 ($p>0.05$; 95\% CI: 0.99--1.06), suggesting a lack of statistical significance in predicting milestone papers.

For the Nobel Prize-winning papers, the odds ratios for $\Delta$ percentile and logarithmic citations were 1.34 ($p<0.001$; 95\% CI: 1.25--1.44) and 5.55 ($p<0.001$; 95\% CI: 5.18--5.95). A 10\% increase in $\Delta$ percentile corresponded to a 1.34 times increase in odds, and a twofold increase in citations associated with a 5.55 times increase. However, the 95\% confidence interval for $D$ percentile includes 1 ($p>0.05$; 95\% CI: 0.96--1.03), suggesting no statistical significance in predicting Nobel Prize-winning papers.

We further extended our analysis to the patent dataset by categorizing government-funded patents as disruptive, based on previous observations that public sector institutions are more likely to focus on far-reaching, foundational research initiatives, while companies tend to concentrate on immediate and practical developments~\cite{funk2017dynamic, li2017applied,fleming2019government}. The result aligns with the observations for scientific papers, showing that $\Delta$ is more effective than $D$ in identifying government-funded patents.(see Supplementary Section 6).

\subsubsection{Simultaneous Disruption}
\label{sec:result_simdis}
\begin{figure}
    \centering
    \includegraphics[width=0.87
    \textwidth]{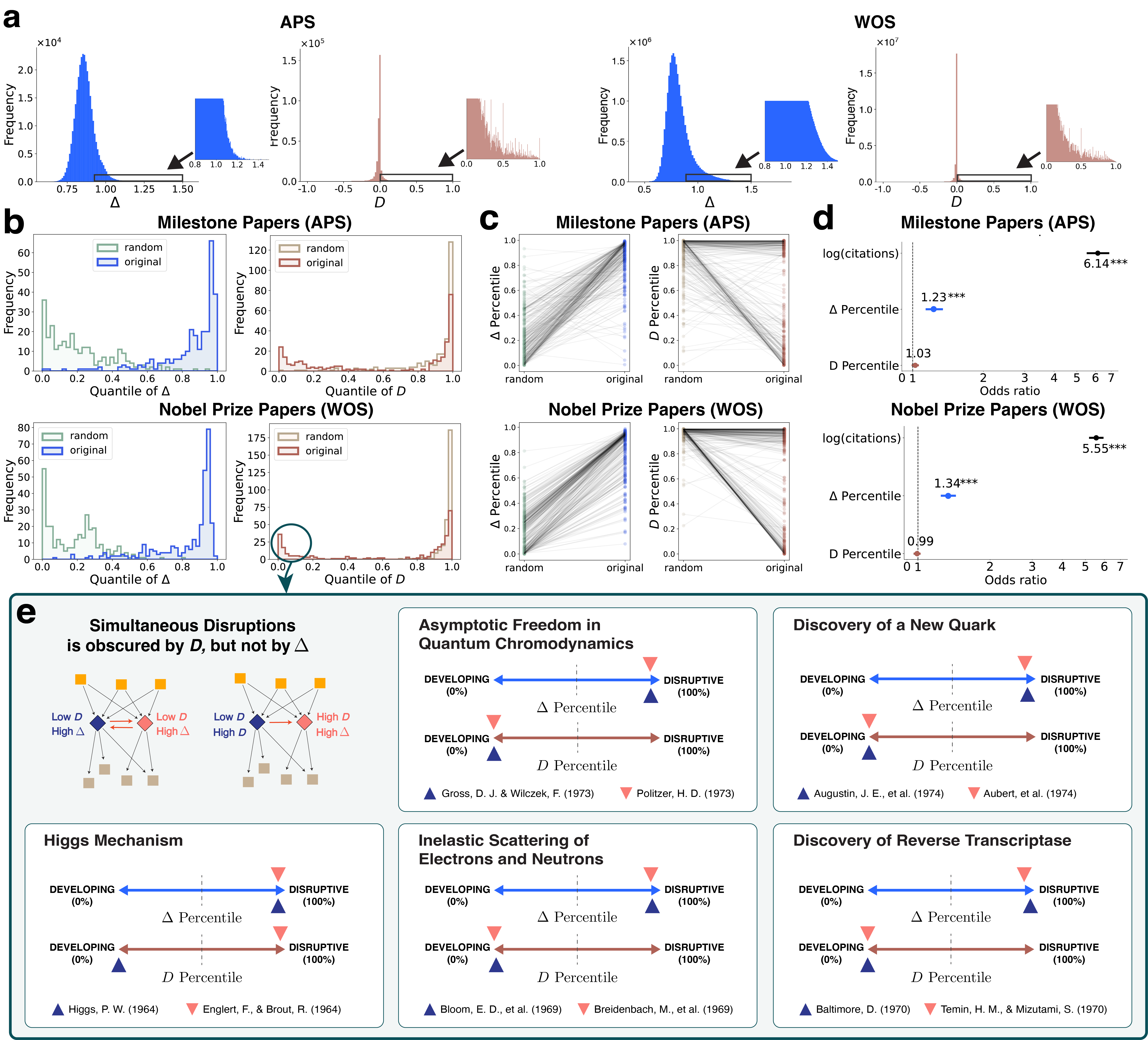}
    \caption{\textbf{Embedding Disruptiveness Measure (EDM) better captures disruptive works as well as simultaneous disruptions, which are obscured by disruptive index.} \textbf{(a)}
    The disruption index $D$ has higher degeneracy than an embedding disruptiveness measure $\Delta$. The disruption $D$ has high degeneracy in specific values such as 1, 0.5, and 0.25 (see Figure~\ref{fig:problems}). \textbf{(b)} The $D$ index of 302 Nobel Prize-winning and 278 milestone papers shows a bimodal distribution, mainly attributed to the failure of $D$ to consider simultaneous disruptions. Remarkably, $\Delta$ successfully rank most of them highly disruptive, eliminating the bimodal distribution. Also, the distribution of two indices in the randomization of the citation network highlights the influence of citations and references on $D$ scores. \textbf{(c)} The change of the percentile of $\Delta$ at the individual paper level varies between the randomized citation network and the original citation network. In contrast, the percentile of $D$ either shifts drastically due to the sensitivity of the index or remains nearly unchanged. \textbf{(d)} Firth's logistic regression shows that $\Delta$ correlates more strongly with the likelihood of papers becoming milestones among 327,021 APS papers or Nobel Prize-winning papers among 23,664,187 WOS papers, with higher and more statistically significant odds ratios than $D$. Error bars represent the 95\% confidence interval. \textbf{(e)} Examples of the papers involved in simultaneous discoveries that were overlooked by the disruption index $D$ but effectively captured by EDM with high $\Delta$ score. The $D$ scores for these papers were positioned around the bottom 1\%, contrasting with a potential ranking higher than the top 5\% if not for the impact of mutual citation links. The downward-pointing triangles on the arrows indicate the percentile of $\Delta$ and $D$ of simultaneous discovery papers.} 
    \label{fig:distribution_disruption}
\end{figure}

If the disruption index is largely confounded by citations and references, then why do some of these landmark papers have such low scores, resulting in a bimodal distribution of $D$? To understand this discrepancy, we examine the top ten papers with the largest difference between $D$ and $\Delta$.

We found that, all ten papers are related to the notable examples of \emph{simultaneous disruption}, where multiple papers independently reached the same conclusion or where authors published the study across two separate publications. The problem of $D$ is that, when there is \emph{even a single citation} linking these simultaneous disruptive papers, $D$ can change, in principle, from the maximum ($D = 1$) to the minimum ($D = -1$), as shown in Figure~\ref{fig:problems}. Due to the discrete nature of classifying future papers into the two types, a single citation edge can turn every descendent paper from one class to the other. Indeed, this seems to be what happened to these papers. 

There are eight simultaneous disruption cases related to these ten papers in total. The details for the five most notable of these cases are illustrated in Figure~\ref{fig:distribution_disruption}e. Let us examine two of these instances here. The first is the story of $J/\psi$ meson, a transformative discovery that challenged the established three-quark model at that time and sparked what is now celebrated as the November Revolution in particle physics. On November 
11, 1974, two independent research teams---one led by Burton Richter and the other led by Samuel Ting---simultaneously announced a discovery of a new particle~\cite{aubert1974experimental,augustin1974discovery}. Both groups published their findings in the same issue of the journal \emph{Physical Review Letters} (PRL) and they both cited each other. As a result of the citations between their papers, compared with the hypothetical case without them, their papers' $D$ dropped from 0.085 (top 4\% percentile) and 0.80 (top 1\%) to $-0.11$ (bottom 1\%) and $-0.64$ (bottom 0\%). By contrast, $\Delta$ correctly captures their collective disruption as 0.99 (top 5\%) and 0.97 (top 7\%). Another case is the Higgs Mechanism. In 1964, the question of why particles possess mass was addressed through the development of a model by François Englert and Robert Brout, as well as independently by Peter Higgs~\cite{englert1964broken,higgs1964broken}. Like the previous case, they published their studies around the same time. However, only the paper by Higgs cites Englert and Brout's paper, resulting in a decrease in the $D$ value of Higgs' paper from 0.28 (top 1.3\%)---the counterfactual case of Higgs not citing Englert \& Brout's paper---to $-0.27$ (bottom 0.1\%). On the other hand, the $D$ value of Englert \& Brout's paper, which did not cite Higgs' paper, is 0.15 (top 2.8\%). Again, the $\Delta$ values for these papers precisely gauge their overall disruption, resulting in scores of 1.002 (top 4.1\%) and 1.005 (top 4.1\%). Other cases of simultaneous disruption shown in the figure are discovery of reverse tanscriptase, asymptotic freedom in quantum chromodynamics, and inelastic scattering of electrons and neutrons. The details of these examples of simultaneous disruption, along with the other three cases not shown in the figure, are represented in Supplementary Information Section 4 and 5. A table listing the top eleven papers with the largest discrepancies between $\Delta$ and $D$ is also included in Supplementary Table~\ref{tab:d_and_delta_nobel} (The inclusion of the eleventh paper is due to its status as a simultaneous discovery pair with the third-ranked paper.)

\subsection{Identification of Simultaneous Discoveries}

\begin{figure}
    \centering
\includegraphics[width=0.95\textwidth]{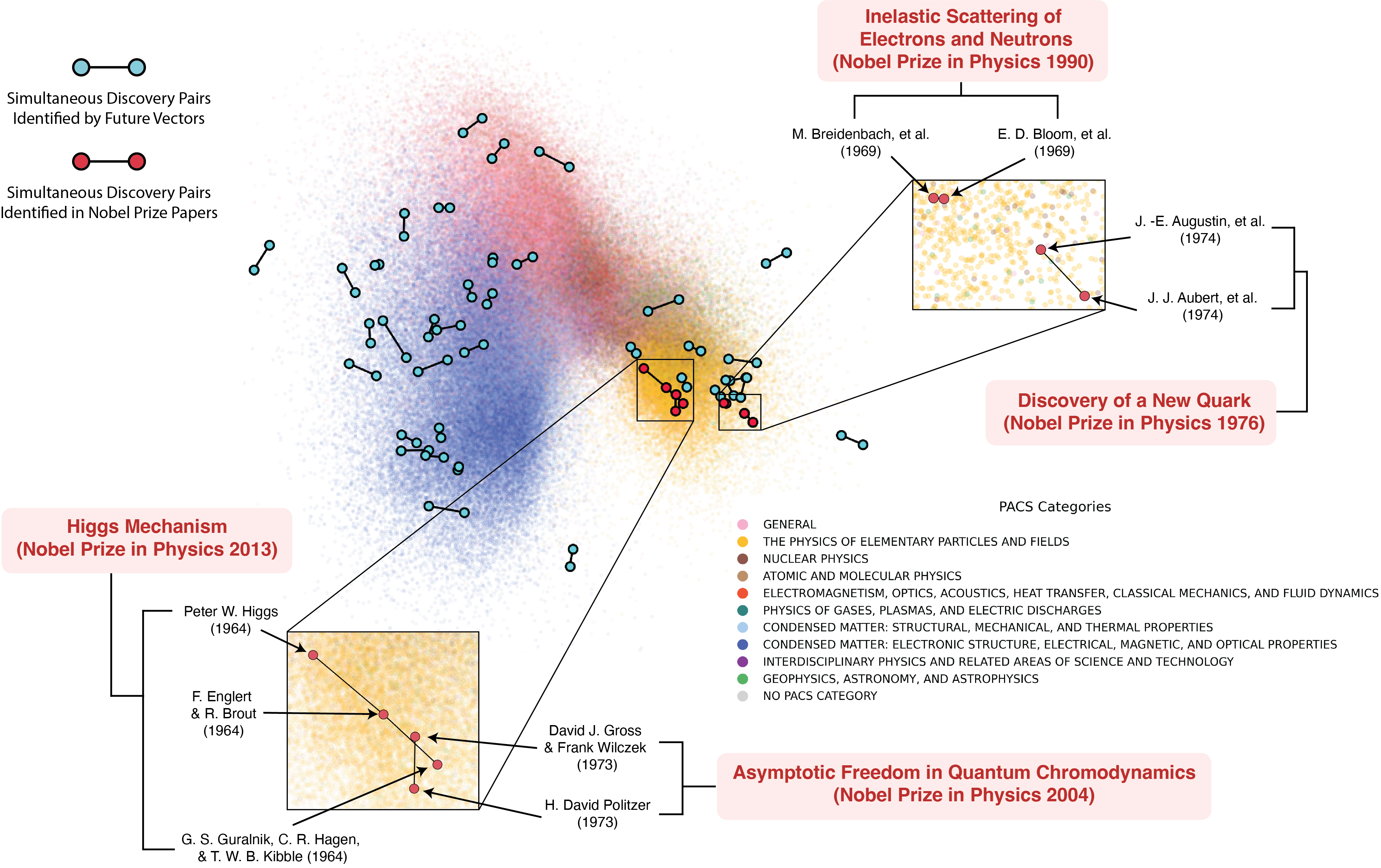}
  \caption{\textbf{Future vectors reveal the simultaneous discoveries.} The PCA plot of the future vectors of 332,518 APS papers. Each paper's future vector is color-coded according to its PACS category. Simultaneous discovery papers identified in the Nobel Prize-winning papers dataset are represented by large red dots, with their closest neighbors connected by black edges. Two papers involved in the discovery of reverse transcriptase are excluded as they are not included in the APS datset. For the Higgs mechanism's simultaneous discovery, we included an additional paper by Guralnik et al.~\cite{guralnik1964global} (see Supplementary Section 4 for details). Notably, the future vectors of these simultaneous discovery papers are the \emph{closest} pairs to each other. Using this feature, we identified 18,417 potential simultaneous discovery papers, among which we annotated 80 papers with more than 300 citations. Of the 80 annotated papers, 64 were identified as simultaneous discovery papers: 34 papers are independent simultaneous discovery papers, where the authors explicitly reference their simultaneous discovery counterparts within the text, while another 30 papers represent collective simultaneous discoveries, where the same authors shared their findings across multiple publications. The 34 independent simultaneous discovery papers are colored with bright teal in the figure. The black edge again shows that the future vectors of these simultaneous paper pairs are closest to each other. Principal components 2 and 3 are shown as principal component 1 primarily captures temporal information.}
    \label{fig:UMAP}
\end{figure}

In the previous section, we demonstrated the advantages offered by EDM in mitigating issues related to degeneracy, locality, and the ability to capture simultaneous disruption.  Notably, we found numerous examples of simultaneous discoveries in the Nobel Prize-winning paper dataset, supporting Robert Merton's argument that simultaneous discovery is not a rare aspect of scientific evolution~\cite{merton1961singletons}.These examples of simultaneous disruption exhibit similarly high EDM scores, indicating that their true significance is effectively captured by our embedding algorithm.

However, EDM is a measure of the discrepancy between future vectors $\mathbf{f}$, which reflect how a paper is used in subsequent works, and past vectors $\mathbf{p}$, which indicate what the paper drew from earlier research. This discrepancy alone does not confirm that the embedding algorithm accurately captures the parallel impact of simultaneous disruptions on future studies. For the embedding algorithm to truly reflect the similarity of their impact, the embedding locations of simultaneous disruptions should be close to one another, as these papers represent the same discoveries and are utilized by the scientific community in similar contexts. Given this, we further investigate whether the future vectors $\mathbf{f}$ of simultaneous discoveries are positioned close to each other in the embedding space. Here, we focus on the proximity of future vectors of simultaneous discovery pairs, reflecting the similarity in how a pair of papers is utilized by subsequent research, following the approach of previous studies~\cite{bikard2020idea,painter2020quantifying,small1973co}. The same analysis on the past vectors is provided in Supplementary Section 10.

We analyzed the proximity of future vectors for simultaneous discovery examples based on cosine distance, as in our approach for calculating $\Delta$, to reduce the influence of citation counts. Due to the computational demands of identifying nearest neighbors across the entire WOS dataset, which includes over 23 million papers, we focused our analysis on the APS dataset. As expected, simultaneous discovery papers identified in the previous section consistently clustered together in the embedding space, showing the closest proximity to each other compared to all other papers. For visualization, we plotted the future vectors of the entire APS dataset using a PCA projection (Figure~\ref{fig:UMAP}).

Given that the simultaneous discovery pairs in our examples consistently appear as the nearest neighbors, we further investigated the potential of our embedding approach to systematically identify such simultaneous discoveries. While previous studies have used overlapping citation patterns to detect simultaneous breakthroughs ~\cite{bikard2020idea,painter2020quantifying,small1973co}, these methods are limited to analyzing one-step citation counts, potentially having similar limitations with the disruption index. In contrast, leveraging future vector proximity enables the identification of simultaneous discoveries by capturing the broader, cumulative influence of papers on subsequent research. 

To test this approach, we first selected paper pairs with the closest future vectors that were published in the same year, yielding 18,417 paper pairs involving 36,834 papers. We then excluded review papers published in Review of Modern Physics (RMP), resulting in 36,743 papers. The dataset of these potential simultaneous discovery paper pairs is publicly available (see Methods for details)

Because of the impracticality of manually verifying each of the 36,743 papers for simultaneous discoveries, we focused on pairs with significant citation impact, those receiving over 300 citations, narrowing the selection to 80 papers. Among these, 64 (80\%) were identified as simultaneous discoveries. Specifically, 34 papers explicitly acknowledged the presence of their independent simultaneous discoveries within their main text or citations. For example, Miyake et al. noted in their work~\cite{miyake2013realizing}, ``Recently, we became aware of similar work carried out by Bloch and co-workers,'' referencing the paper by Aidelsburger et al.~\cite{aidelsburger2013realization}, which exhibited the closest future vectors to Miyake’s paper. These independent simultaneous discovery paper pairs identified by our embedding approach are illustrated in Fig.~\ref{fig:UMAP}. Other 30 papers were identified as instances of collective simultaneous discoveries, where the same authors disseminated their findings across multiple publications. The dataset for these simultaneous discovery annotations is available (see Data Availability).

\section{Discussion}

In our study, we highlight key limitations of the disruption index and introduce a neural embedding framework that provides a more robust measure of scientific disruptiveness, with further applications in identifying simultaneous discoveries. This method not only facilitates the discernment of the true influence of disruptive papers, overcoming limitations inherent in existing metrics but also provides a means of identifying simultaneous discoveries.

While our study makes significant strides in overcoming the existing problems of the disruption index, our metric has notable limitations. First, measuring temporal changes of disruptiveness is challenging with our measure. Because our metric consider the entire structure of the citation network during the training process, the impact of a paper at specific points in time can only be quantified by retraining the model using citation data up to that period, which can be computationally demanding. Second, our method struggles to capture the disruptive impact of papers with few citations or references. Papers with no citation or no reference are excluded from the random walk sequences, meaning they are entirely left out of the training process. Third, the metric primarily focuses on citation relationships, which, while meaningful, do not always fully capture the broader influence of a paper. Contributions that impact the field in ways not reflected through citations, such as informal dissemination or cases where review articles replace the original citation (see Supplementary Section 3), may not be adequately represented.

Despite these limitations, our metric represents a substantial improvement in capturing the multifaceted nature of disruptiveness in scientific research, by addressing significant limitations in existing indices through the application of neural embedding methods. The demonstrated reduction in degeneracy, comprehensive consideration of the entire network structure, and resistance to misleading results in the face of collective and simultaneous disruption examples underscore the efficacy of our proposed metrics. Given the centrality of disruptiveness in characterizing the progression of scientific knowledge, our metric offers valuable insights into the diverse mechanisms driving scientific evolution. Furthermore, its ability to identify the simultaneous disruption sets opens doors for investigating the drivers of innovation and recognition dynamics of scientific works in the scientific community.

\printbibliography[ title={References}]

\begin{refsection}
\section{Methods}

 \subsection{Dataset}
\label{sec:method_dataset}

\subsubsection{WOS Papers}

We use a dataset from WOS (Web of Science) including papers released from 1960 to 2019, which amount to 54,945,692 papers connected by 1,077,709,818 citations. We compute both the disruption index $D$ and our new index $\Delta$ using this full data set. After calculating these indices, we exclude the papers between 1960 and 1962 as well as those between 2017 and 2019 to alleviate the boundary problem---biased computation of the indices due to missing data before and after the time period captured in the data. Furthermore, after calculating the disruption index, we do not consider papers with fewer than five citations and no references to exclude outliers and ensure that every paper belongs to the giant connected component. We also narrow our focus down to only research-related document types (e.g., `Review,' `Article,' `Letter,' and `Proceedings Paper').  After these considerations, the total number of papers we include in our study is 23,664,187.

\subsubsection{APS Papers}
The APS dataset covers articles published in APS journals, comprising 644,022 papers published from 1893 to 2019 with 8,323,911 citations. Similar to our approach with the WOS data set, we calculated disruption scores and selected papers with a minimum of five citations and at least one reference. Filtering papers with less than five citations or with no reference remains the papers published after 1913, so we only further exclude the papers published after 2017, resulting in a final tally of 327,021 papers. For the identification of simultaneous discovery pairs in Result, we include papers without references since we solely utilize future vectors, which is based on only citation information, bringing the total to 332,518 papers.

\subsubsection{Nobel Prize-Winning Papers}
\label{subsec:nobel-prize-winning}

In our study, we identified 302 Nobel Prize-winning papers, sourced from two comprehensive studies. The first study compiled 874 prize-winning papers published between 1900 and 2016 across three major fields: 283 in Physics, 259 in Chemistry, and 332 in Medicine. This dataset was constructed by analyzing references cited in Nobel lectures, with additional cross-verification using sources such as Wikipedia, personal homepages, and existing academic studies to ensure the accuracy and completeness of the identified papers~\cite{DVN/6NJ5RN_2018}. The second study focused on Nobel Prize-related publications from 1995 to 2017, selecting one significant paper per laureate based on citation count and main authorship. These key papers were identified from the Advanced Information section on the official Nobel website, resulting in a selection of 174 papers~\cite{ioannidis2020work}. For this source, we identified the papers using their Digital Object Identifiers (DOIs). From these combined sources, we compiled a final dataset of 302 unique Nobel Prize-winning papers for our analysis.

\subsubsection{Milestone Papers}

We manually collected the milestone paper datasets from each journal website. As a result, we collected 79 papers from 125 Years of the American Physical Society Journals milestones~\cite{APS125Years}, 87 papers from a PRL retrospective milestones~\cite{prl_milestone}, 24 papers from Physical Review A 50th Anniversary Milestones~\cite{pra_milestone}, 55 papers from Physical Review B 50th Anniversary Milestones~\cite{prb_milestone}, 39 papers from Physical Review C 50th Anniversary Milestones~\cite{prc_milestone}, 85 papers from Physical Review D 50th Anniversary Milestones~\cite{prd_milestone}, and 22 papers from Physical Review E 50th Anniversary Milestones~\cite{pre_milestone}. Due to overlaps between milestone selections in 125 Years of the American Physical Society Journals and other journals, the collection totals 300 milestone papers, with 278 papers included in the datasets used in the analysis.

\subsubsection{Patents}

We gathered 7,387,609 patents granted by the United States Patent and Trademark Office (USPTO) from 1976 to 2020, sourced from PatentsView~\cite{uspto_patentsview}. We computed a disruption index using the entire citation network and refined the dataset by excluding patents with fewer than five citations or only one reference, as well as non-utility patents, such as design patents and defensive publications. Additionally, we filtered out patents issued after 2016 and before 1979, with the same reason of excluding the first and the last three years of WOS papers and APS papers, resulting in 2,653,873 patents. PatentsView also offers government interest patents granted by the USPTO for inventions funded, at least in part, by a federal research grant or government contract. There are 177,742 government interest patents in the datasets, and among them, 56,243 patents are included in the dataset we used in the analysis.

\subsection{Embedding Disruptiveness Measure}
\label{sec:method_edm}

To calculate the disruptive features of papers, we first imagine two separate representations of each paper, which we call ``past vector'' and ``future vector.'' We assume that having good representations means that the past vector can effectively predict the previous nodes in random walk trajectories (i.e., the antecedent papers that the given paper cites), while the future vector can predict the next nodes (i.e., the descendant papers that cite the given paper). Borrowing ideas from statistical language models and graph embeddings, we further assume that this implies the past vector aligns closely with the antecedent papers, while the future vector aligns with the descendant papers~\cite{grover2016node2vec,mikolov2013distributed, perozzi2014deepwalk, tang2015line,}.

In this framework, if a paper is disruptive, its descendants (those citing it) rely less on its antecedents (the papers it cites), as the given paper makes the antecedents obsolete or not relevant to the descendent papers~\cite{funk2017dynamic}. This creates a clear separation between the two, causing the representations of the antecedent and descendant papers to move farther apart. Consequently, the past and future vectors become increasingly distant, indicating the paper’s disruptiveness and suggesting that future research is diverging from the traditional context set by the antecedents. Therefore, the disruptiveness of a paper can be quantified as the distance between its future and past vectors.

To formalize this concept, we first frame the process of learning features of papers as an optimization problem. Let $G = (V, E)$ represents the citation network, where $V$ and $E$ indicate the set of papers and citation links, respectively. For any given paper $u \in V$, we define $D_c(u) \subset V$ as the collection of descendant papers that cite paper $u$ within $c$ citation steps, and $A_c(u) \subset V$ as the collection of antecedent papers that are cited by the paper $u$ within $c$ citation steps.

Then, inspired by the skip-gram model, which is a statistical language model designed to predict surrounding words in a sentence~\cite{mikolov2013distributed}, we define the objective of our model as 

\begin{equation}\label{eqn:objective_randomwalk_method}
\mathcal{J} = \sum_{s\in V} \sum_{r=1}^{R} \sum_{t=1}^{T - c} \sum_{w=1}^{c} \log \text{Pr}(v_{t+w}^{(r, s)} | v_t^{(r, s)}),
\end{equation}
which is the sum of the log-probabilities of observing $v_{t+w}^{(r, s)}$ as a cited paper within $c$ citation steps of $v_t^{(r, s)}$, across $R$ random walks of length $T$ initiated from each paper $s \in V$. In this context, $v_{t}^{(r, s)}$ refers to the paper at position $t$ in the $r$-th random walk starting from $s$, while $v_{t+w}^{(r, s)}$ is the paper cited by $v_{t}^{(r, s)}$ through $w$ citation steps.

Random walks in directed networks tend to visit nodes with a large in-degree $\kappa^{\text{in}}$. It is known that the probability of visiting each node is approximately proportional to its in-degree ~\cite{fortunato2007random}. Keeping this in mind, let us simplify Equation (\ref{eqn:objective_randomwalk_method}). Equation (\ref{eqn:objective_randomwalk_method}) involves $R$ random walks of length $T$, represented by the second and third summation. When $R$ and $T$ are big enough, a random node $u$ appears with frequency approximately proportional to its in-degree $\kappa^{\text{in}}_u$. Thus, the two summations are simplified by $\kappa^{\text{in}}_u$,  and Equation (\ref{eqn:objective_randomwalk_method}) can be approximated as 
\begin{equation}\label{eqn:objective}
\mathcal{J} \approx \sum_{u \in V} \sum_{v \in A_c(u)} \kappa_u^{\text{in}}\log \text{Pr}(v | u),
\end{equation}
where $\text{Pr} (v|u)$ is the conditional probability to observe $v$ in $A_c(u)$ when the paper $u$ is given. In our model,  $\text{Pr} (v|u)$ is calculated as the softmax function of the similarity between $\mathbf{f}_v$, which is the future vector of the antecedent works $v$, and $\mathbf{p}_u$, which is the past vector of the target work $u$:

\[
    \text{Pr} (v|u) = \frac{ \exp({\mathbf{f}_v \cdot \mathbf{p}_u})} {Z_u},
\]
where $Z_u = \sum_{v' \neq u} \exp{(\mathbf{f}_{v'}\cdot\mathbf{p}_u)}$ (In fact, the function $\text{Pr}(v|u)$ is biased towards the noise distribution due to the use of negative sampling in the method~\cite{kojaku2021residual2vec}.  A derivation of the same conclusion, accounting for the bias, can be found in Supplementary Section 9.)

The difference from the original skip-gram formulation is that our model restricts the prediction window to only the ``left" context in the citation network~\cite{song2018directional}. This means that we focus on predicting the antecedent papers by using the past vector of the given paper and the future vectors of its antecedent papers, rather than employing a symmetric window around the target. This design choice reflects our initial assumption that a paper's influence is captured through its ``future vector," which becomes distinct from its ``past vector" when the paper is disruptive.

In the following, we will show that the cosine distance between the future vector and the past vector of the given paper $i$ can approximate the lack of reliance of $D_c(i)$, the descendent works of $i$, on the $A_c(i)$, the antecedent works of $i$.

In Equation (\ref{eqn:objective}), the terms related to the given paper $i$ are
\begin{align*}
  \sum_{j\in A_c(i)} \kappa_i^{\text{in}}\log \text{Pr}(j | i) + \sum_{k \in D_c(i)} \kappa_{k}^{\text{in}}\log \text{Pr}(i | k) & =  \sum_{j\in A_c(i)}  \kappa_i^{\text{in}}\log \frac{ \exp({\mathbf{f}_j \cdot \mathbf{p}_i})} {Z_i} + \sum_{k\in D_c(i)} \kappa_{k}^{\text{in}}\log \frac{ \exp({\mathbf{f}_i \cdot \mathbf{p}_k})} {Z_{k}} \\
  & =  \sum_{j\in A_c(i)} \kappa_i^{\text{in}} \left({\mathbf{f}_j \cdot \mathbf{p}_i} - \log {Z_i}\right) +  \sum_{k\in D_c(i)}\kappa_{k}^{\text{in}}\left(\mathbf{f}_i \cdot \mathbf{p}_k -\log {Z_{k}}\right).\\
 \stepcounter{equation}\tag{\theequation}\label{myeq1}
\end{align*}

We assume that the network is acyclic, and either $D_c(i)$ or $A_c(i)$ does not include $i$, which is mostly the case for citation networks. Then, the question to find a vector $\mathbf{f}_i$ that maximizes the Equation (\ref{myeq1}) becomes maximizing the following function: 
\[
     \sum_{k \in D_c(i)} \kappa_k^{\text{in}} (\mathbf{f}_i \cdot \mathbf{p}_k - \log Z_k),
\]

where $Z_k = \sum_{v \neq k} \exp{(\mathbf{f}_{v} \cdot \mathbf{p}_k)} = \exp(\mathbf{f}_i \cdot \mathbf{p}_k) +C.$ Note that if we think about a function 
\[
F(s) = s - \log(  e^s +C),
\]
then the gradient of $F(s)$ is positive ($C$ is a positive constant):
\[
    F'(s) = 1 - \frac{e^s}{ e^s +C}
\]
In other words, $F(s)$ is a monotonically increasing function. It is not bounded and would keep increasing with $s$. In practice, due to the noise in the data as well as the stochastic nature of the optimization algorithm (negative sampling and stochastic gradient descent) the norm of every vector is bounded. Thus, if we consider a single term associated with a specific $k$, we can see that $\mathbf{f}_i$, if its norm is fixed, should be aligned with $\mathbf{p}_k$ to maximize the objective function. 

Then, let us assume that set $D_c(i)$ for most papers capture a fairly homogeneous set of publications that are similar to each other. Then, we can think about $\mathbf{p}_k$ as a set of vectors that are distributed around a certain mean direction $\mathbf{u}_j$. Then we can write $\mathbf{p}_k = \mathbf{u}_i + \epsilon_{ik},$ where $\epsilon_{ik}$ is a small perturbation vector uniformly distributed on a hypersphere $(\frac{\sum_{k \in D_c(i)} \mathbf{p}_k }{|D_c(i)|} \simeq \mathbf{u}_i).$ Then,

\begin{align*}
     \sum_{k \in D_c(i)} \kappa_k^{\text{in}} (\mathbf{f}_i \cdot \mathbf{p}_k - \log Z_k) &= \sum_{k\in D_c(i)}\kappa_{k}^{\text{in}} \left[ \mathbf{f}_i \cdot (\mathbf{u}_i +  \mathbf{\epsilon}_{ik}) - \log ( \exp(\mathbf{f}_i \cdot (\mathbf{u}_i +  \mathbf{\epsilon}_{ik})) + C ) \right] \\
&= \sum_{k\in D_c(i)}\kappa_{k}^{\text{in}} \left[ \mathbf{f}_i \cdot \mathbf{u}_i +  \mathbf{f}_i \cdot \mathbf{\epsilon}_{ik} - \log ( \exp(\mathbf{f}_i \cdot \mathbf{u}_i) \exp(\mathbf{f}_i \cdot \mathbf{\epsilon}_{ik}) + C ) \right] \\
&\simeq  \sum_{k\in D_c(i)}\kappa_{k}^{\text{in}} \left[ \mathbf{f}_i \cdot \mathbf{u}_i - \log ( \exp(\mathbf{f}_i \cdot \mathbf{u}_i ) + C) \right] + \sum_{k\in D_c(i)}\kappa_{k}^{\text{in}} \mathbf{f}_i \cdot \mathbf{\epsilon}_{ik} \\
&\simeq \sum_{k\in D_c(i)} \kappa_{k}^{\text{in}}\left[ \mathbf{f}_i \cdot \mathbf{u}_i - \log ( \exp(\mathbf{f}_i \cdot \mathbf{u}_i ) + C) \right],
\end{align*}

where we assume $\exp(\mathbf{f}_i \cdot \epsilon_{ik}) \simeq 1$ and $\sum_{k \in D_c(i)} \kappa_k^{\text{in}} \mathbf{f}_i \cdot \epsilon_{ik} \simeq 0.$ This function is again a monotonically increasing function of $\mathbf{f}_i \cdot \mathbf{u}_i.$ If $|\mathbf{f}_i|$ is fixed, $\mathbf{f}_i$ should be aligned with $\mathbf{u}_i$ to maximize the function.

Now let's consider the past vectors of $i$, $\mathbf{p}_i.$ The objective function that we need to maximize in terms of $\mathbf{p}_i$ is 

\begin{equation}
    \sum_{j \in A_c(i)} \kappa_i^{\text{in}} \left( {\mathbf{f}_j \cdot \mathbf{p}_i} - \log {Z_i} \right)  =    \sum_{j \in A_c(i)} \left(\kappa_i^{\text{in}}  \mathbf{f}_j \cdot \mathbf{p}_i \right) - \kappa_i^{\text{in}}  
 |A_c(i)|\log {Z_i}   , \label{eq:objective_pi}
\end{equation}

where \( Z_i = \sum_{v \neq i} \exp(\mathbf{f}_v \cdot \mathbf{p}_i) \). Still Equation~(\ref{eq:objective_pi}) is a monotonically increasing function of \(\mathbf{v}_i \cdot \mathbf{p}_i\), where \(\mathbf{v}_i\) is approximately the mean future vector of the antecedent papers, i.e., 
$\mathbf{v}_i \simeq \frac{\sum_{j \in A_c(i)} \mathbf{f}_j}{|A_c(i)|}$
(see Supplementary Section 8 for details). Therefore, for a given norm \(\Vert \mathbf{p}_i \Vert\), \(\mathbf{p}_i\) should be as closely aligned with \(\mathbf{v}_i\) as possible to maximize Equation~(\ref{eq:objective_pi}).

Then by calculating the cosine distance between $\mathbf{f}_i$ and $\mathbf{p}_i$, we can approximate the cosine distance between the average future vector of antecedent papers $\frac{\sum_{j \in A_c(i)}\mathbf{f}_j}{|A_c(i)|}$ and the average past vector of descendent papers, $\frac{\sum_{k \in D_c(i)}  \mathbf{p}_k}{|D_c(i)|}$.  The cosine distance between these two average vectors increases as $ \sum_{j \in A_c(i)}\sum_{k \in D_c(i)} \log \text{Pr}(j|k)$ decreases, which reflects the lack of connections between descendent works and antecedent works as:

\begin{align*}
1- \frac{\sum_{j\in A_c(i)}\mathbf{f}_j}{|\sum_{j\in A_c(i)}\mathbf{f}_j|} \cdot \frac{\sum_{k \in D_c(i)} \mathbf{p}_k  }{|\sum_{k \in D_c(i)} \mathbf{p}_k|}  &  = 1- \frac{\sum_{j \in A_c(i)}\sum_{k \in D_c(i)} \mathbf{f}_j \cdot\mathbf{p}_k}{|\sum_{j \in A_c(i)} \mathbf{f}_j||\sum_{k \in D_c(i)} \mathbf{p}_k|  }\\
 &=   1 -\frac{\sum_{j \in A_c(i)}\sum_{k \in D_c(i)}  \left(\log \text{Pr}(j|k) + \log Z_{k} \right)}{|\sum_{j\in A_c(i)}\mathbf{f}_j||\sum_{k\in D_c(i)} \mathbf{p}_k|  }.
\end{align*}

Taken together, the cosine distance between $\mathbf{f}_i$ and $\mathbf{p}_i$ increases when the descendent works of the paper $i$ less rely on its antecedent works. This aligns with the idea of the disruption index, which seeks to capture the degree to which a paper disrupts existing knowledge by introducing new ideas that make future works less dependent on prior literature. Therefore, we define an Embedding Disruption Measure (EDM) index of paper $i$ as 

\[
    \Delta_i =1-\frac{\mathbf{f_i}\cdot \mathbf{p_i}}{\| \mathbf{f_i}\| \|\mathbf{p_i}\|}.
\]

Alternatively, the angle between the two vectors $\mathbf{f}$ and $\mathbf{p}$, or ``geodesic'' distance, calculated as  
$\arccos \Big(
    \frac{\sum_{j\in A_c(i)}\mathbf{f}_j}{|\sum_{j\in A_c(i)}\mathbf{f}_j|}
    \cdot 
    \frac{\sum_{k \in D_c(i)} \mathbf{p}_k}{|\sum_{k \in D_c(i)} \mathbf{p}_k|}\Big)
$,
could also serve as a metric for quantifying the relationship between future and past vectors. Either choice will not alter our analysis, because we focus on the percentile-based evaluation of the cosine distance metric, which makes the specific choice of angle or cosine distance immaterial for our conclusions.

In the main result, we present the outcomes of a model trained with an embedding dimension of $d = 100$ and a window size of $w = 5$. We also observe the robustness of the result across different parameters as shown in Supplementary Figure~\ref{fig:sup_robustness}.

\subsection{Randomized Citation Network}
\label{sec:method_null_model}
To isolate the contribution of disruption measures beyond citation counts and reference patterns, we constructed a null model by creating a randomized citation network. The process began by identifying all pairs where papers published in a specific year $i$ cite the papers published in another year $j$. With these pairs, we created a list of papers published in $i$ and a corresponding list of year $j$ papers that were cited by those published in year $i$. Once these lists were established, the set of cited papers from year $j$ was shuffled, effectively randomizing the connections while preserving the original number of citations and references, as well as the temporal relationships between the citing and cited papers. After shuffling, the randomized list was used to rewire the citation connections, replacing the original links in the network. This randomized network, created by reshuffling the cited papers, serves as the null model, providing a baseline for comparing the original disruption measures and identifying the unique aspects of disruption that are not simply artifacts of citation and reference counts.

\subsection{Firth's logistic regression}

Given that the Nobel Prize-winning papers and milestone papers constitute a very small fraction of the entire paper dataset, standard logistic regression may produce biased and unreliable estimates. Therefore, for estimating the association between $\Delta$, $D$, and the citation counts of papers with the probability of a paper being identified as Nobel Prize-winning and milestone, we used Firth's logistic regression, a bias-reduction technique that generates more reliable estimates, particularly in cases of small sample sizes or imbalanced data~\cite{firth1993bias}. We implemented Firth's logistic regression using the \texttt{logistf} package in R version 3.6.3~\cite{firth_package}.

\printbibliography[ resetnumbers=false,  title={Methods References}]
\end{refsection}

\section*{Funding}
This work was supported by the Air Force Office of Scientific Research under Grant No. FA9550-19-1-0391 and the National Science Foundation under Grant No. 2404109.

\section*{Acknowledgments}

We acknowledge NVIDIA Corporation for their GPU resources and express our gratitude to Kaleb Smith from the NVIDIA SAE-Higher Education Research team for his help of GPU optimization. We also thank to Donghae Seo for reviewing our analysis of the contributions of the Nobel Prize-winning paper in Physics and our mathematical argument. Lastly, we would like to thank Alessandro Flammini, Filippo Menczer, James Evans, Lili Miao, Nadav Kunievsky, Hongbo Fang, Jisung Yoon, Taekho You, Damin Lee, Isabel Constantino, and Govind Gandhi for their valuable comments.

\section*{Contribution}

M.J.K., S.K., and Y.Y.A. conceived the project and designed the analysis. S.K. developed the initial code for the directional skip-gram, which M.J.K. modified and extended into a package for calculating embedding disruptiveness. M.J.K. conducted the data analysis, ran the models, created the visualizations, and drafted the manuscript under Y.Y.A.'s guidance. All authors collaboratively reviewed and revised the manuscript.

\section*{Competing interests}
The authors declare no competing interests.

\section*{Code Availability}
\sloppy
The code for reproducing the results presented in this paper is available at \url{https://github.com/yy/embedding-disruptiveness}. Additionally, the core implementation for calculating the Embedding Disruptiveness Measure (EDM) is provided as a Python package, \texttt{embedding-disruptiveness}. For more details and installation instructions, visit the package's page: \url{https://test.pypi.org/project/embedding-disruptiveness/0.1.2/}.

\section*{Data Availability}
Our study relies on data sourced from the American Physical Society (APS), Web of Science (WoS), and PatentsView. APS dataset can be obtained by contacting APS. The APS dataset is accessible upon request via the APS website at \url{https://journals.aps.org/datasets}. WOS data are not publicly accessible and were utilized under licensing agreements with their respective publisher. The dataset of Nobel Prize-winning papers is available from two referenced publications~\cite{DVN/6NJ5RN_2018,ioannidis2020work}. Patents dataset can be obtained in PatentsView website. Additionally, the data generated and analyzed in this study---including the calculated the disruption index, embedding disruptiveness measure, Digital Object Identifiers (DOIs) for APS dataset, WOS code for WOS dataset, milestone collection papers, and identified simultaneous discoveries within the APS dataset---are publicly available at \url{https://figshare.com/s/c6b0303cfeb549d742c0}.

\fussy

\begin{refsection}

\newpage
\makeatletter
\def\fnum@figure{\figurename\thefigure}
\makeatother
\setcounter{figure}{0}    
\renewcommand{\figurename}{Supplementary Figure }

\section*{\centering Supplementary Information}
\begin{center}
    {\Large \textbf{Uncovering simultaneous breakthroughs with a robust measure of disruptiveness}} \\[0.5cm]
    \normalsize
    Munjung Kim$^{1}$, Sadamori Kojaku$^{2}$, and Yong-Yeol Ahn$^{1}$ \\[0.3cm]
    $^{1}$Center for Complex Networks and Systems Research, Luddy School of Informatics, Computing, and Engineering,
Indiana University, Bloomington \\
    $^{2}$School of Systems Science and Industrial
Engineering, Binghamton University, State University of New York \\
\end{center}

\label{suppinfo}
\subsection*{1. Analysis on Variation of Disruption Index}
\label{suppinfo:variation}
\begin{figure}[htbp]
    \centering
    \includegraphics[width=1\textwidth]{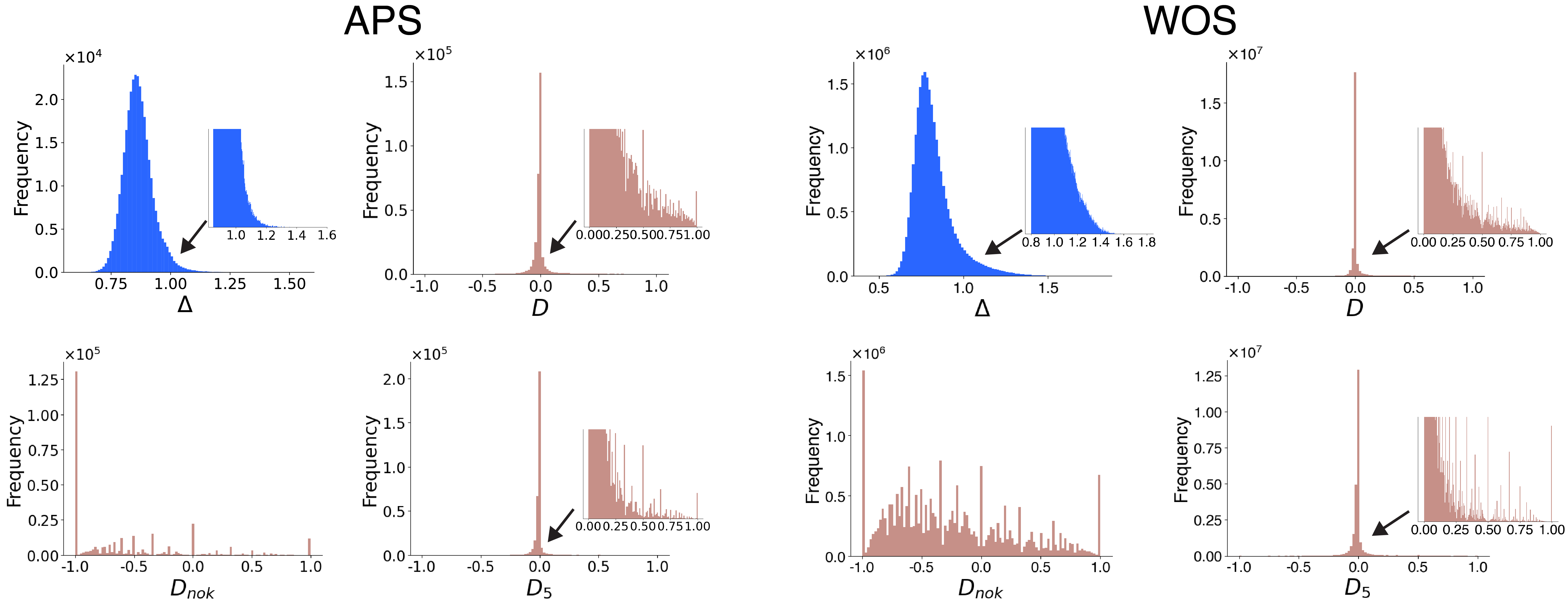}
    \caption{  
    \textbf{Disruption index $D$ and its variants have high degeneracy while our new disruption index $\Delta$ does not.} Two variants of $D$ are explored: $\Dnok$, where the dominant influence of the $n_k$ term is mitigated; and $D_{5}$, the disruption index considering only citations occurring 5 years after the publication of the focal paper. Both indices of APS papers ($n =327,021$) and WOS papers ($n=23,664,187$) revealed higher degeneracy than the original index $D$.  }
\label{fig:sup_distribution_disruption}
\end{figure}

We further explore two common variants of the $D$ index: $\Dnok$ and $D_{5}$. In $\Dnok$, the $n_k$ term in $D$ is dropped because of its dominant influence on the index, which tends to shrink the index toward zero~\cite{wu2019solo}. We found that the distribution of $\Dnok$  does not exhibit a pronounced concentration near zero, as observed in the original $D$ distribution, which is attributed to the omission of the $n_k$ term. However, it does display much higher degeneracy. $D_5$ is the disruption index where only the citations 5 years after the focal paper is published are considered~\cite{park2023papers}. The $D_5$ index still shows a high concentration near the value zero, with higher degeneracy than the index $D$.

\subsection*{2. Analysis of the Locality of the Measures}
\label{suppinfo:additional_disruption}
\begin{figure}[htbp]
    \centering
    \includegraphics[width=0.9\textwidth]{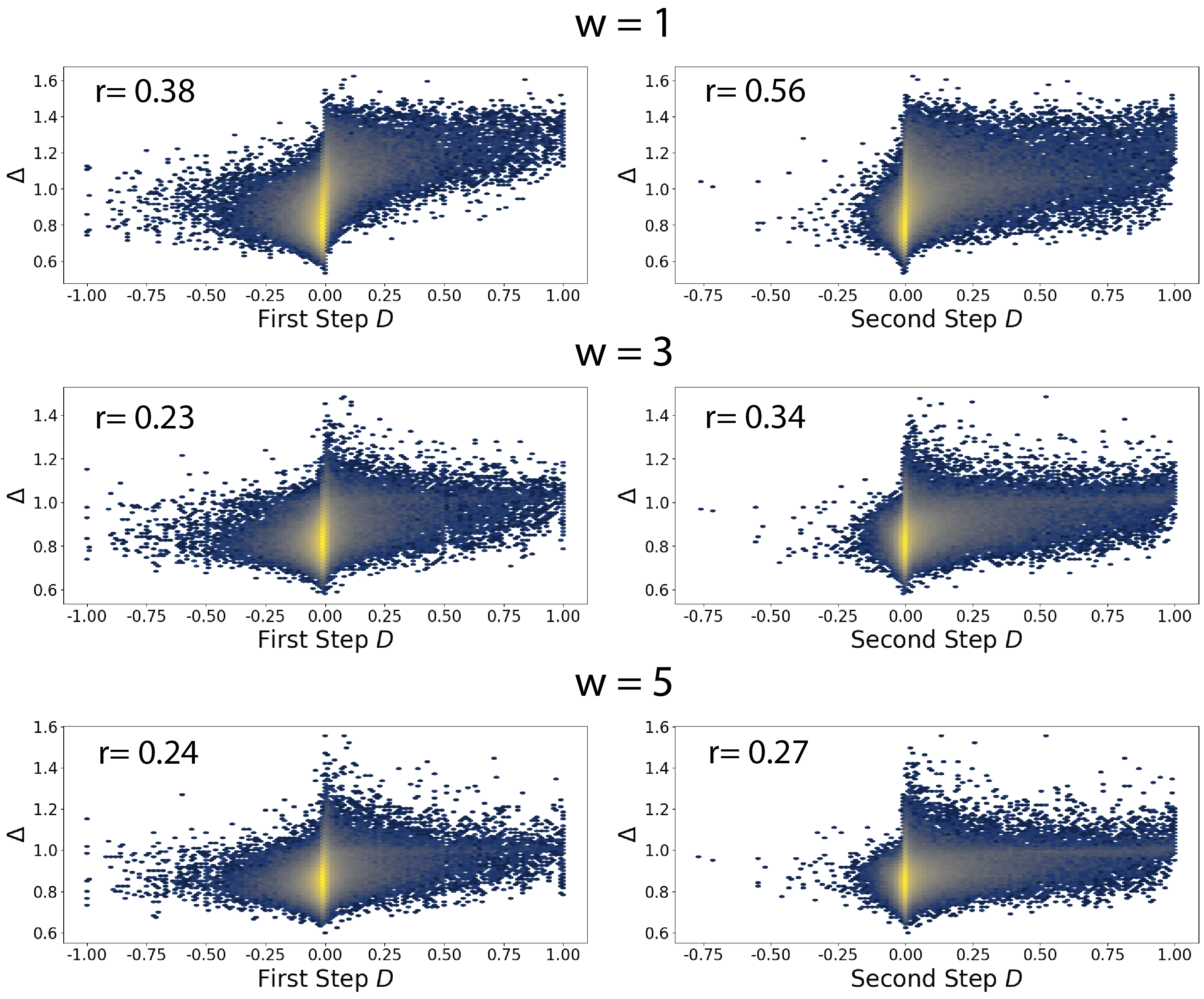}
    \caption{ \textbf{The Embedding Disruptiveness Measure (EDM) addresses locality issues inherent in the Disruption index.} The Spearman's rank correlation between $D$ and $\Delta$ increases ($n = 327,021$), when the second citation step is considered in the calculation of $D$. This highlights the capacity of $\Delta$ to encompass a more extensive spectrum of information, surpassing the constraints associated with relying solely on a single citation. }
\label{fig:distribution_disruption1}
\end{figure}

We examined the locality of the measures by computing the $D$ index accounting for two citation steps and comparing the resulting values with $\Delta$. In this approach, the papers cited by the focal paper through two-step citation steps are regarded as antecedent papers, while the papers that cite the focal paper through two-step citation steps are considered descendant papers.
Given the substantial computational load required to calculate the $D$ index over two citation steps, we limited our analysis to the APS dataset. The comparative results shown in Supplementary Figure~\ref{fig:distribution_disruption1}, indicates that Spearman's correlation for $\Delta$ is always higher when compared with the two-step $D$ index than when compared with the one-step $D$ index, regardless of the model's window size. These increased correlations suggests that the $\Delta$ index captures a broader spectrum of information beyond a single citation.

\subsection*{3. Analysis on Review paper}
\label{supplementary:subsection_review}
\begin{figure}[htbp]
    \centering
    \includegraphics[width=0.8\textwidth]{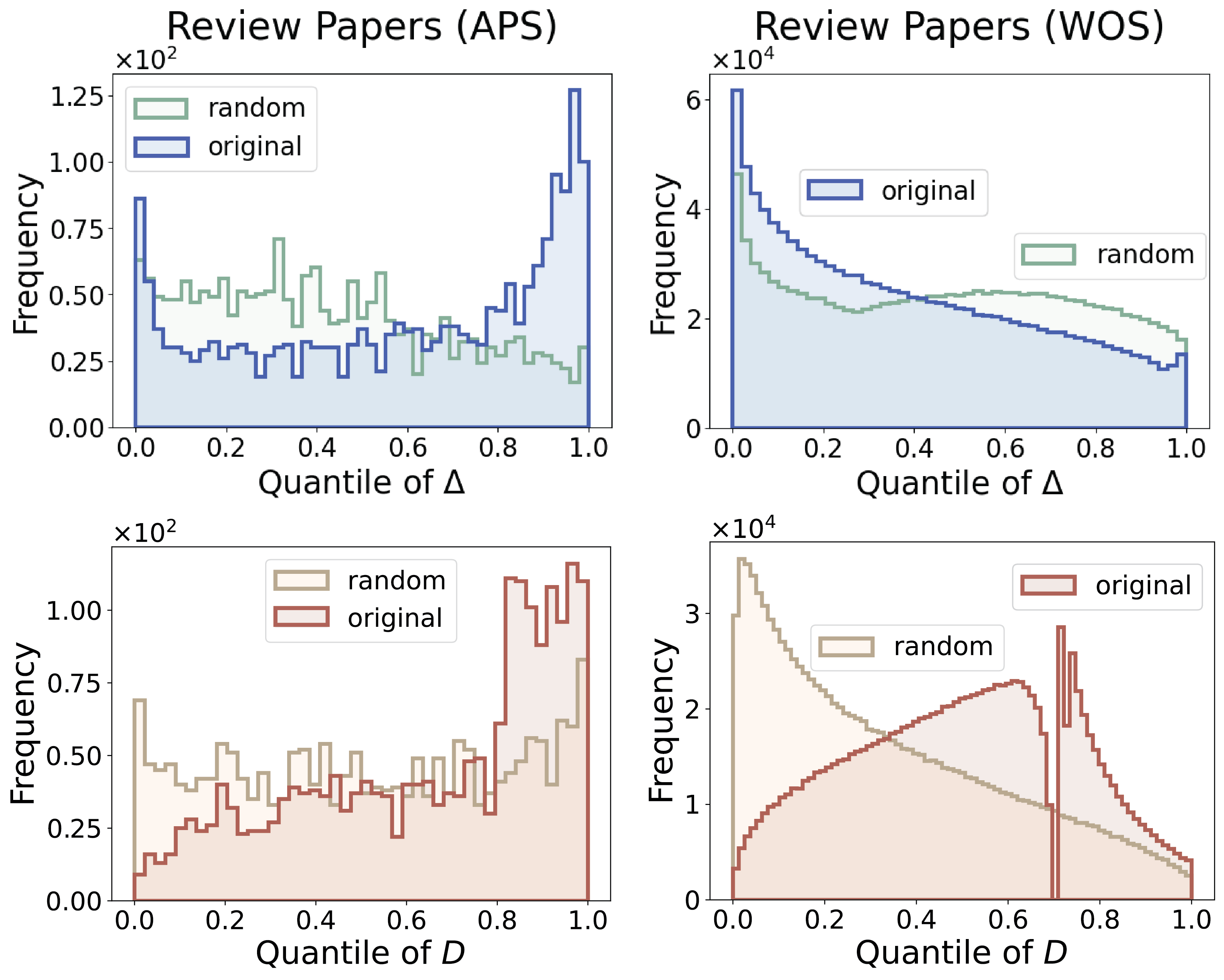}
    \caption{  
    \textbf{Analysis of Disruption Indexes Applied to Review Papers.} Analysis of review articles in the WOS dataset ($n=1,166,514$) reveals that $\Delta$ scores are distributed lower than the baseline $\Delta$ scores measured in the random network, showing the developing characteristics of review articles. However, $D$ scores are distributed higher than the baseline $D$ scores, implying that their disruptiveness is greater than the random network baseline. The Review of Modern Physics papers in APS dataset ($n=2,014$) , a review journal from APS, both indices indicate disruptive contributions exceeding the baseline scores. This trend may be attributed to the journal's high impact (IF 44.1). Review articles of significant influence often supplant previous work references and pave the way for new directions in future studies. }
\label{fig:sup_distribution_disruption_review}
\end{figure}

Review papers are academic papers that summarize and synthesize existing research. They are typically categorized as developmental papers, as they do not usually present novel innovations but `review' what had happened in the field. In some cases, however, review papers can go beyond mere summarization; They can clarify the controversies in the emergent fields or suggest new directions for future research, having a transformative impact on the related fields~\cite{palmatier2018review,paul2021writing}. Moreover, when a field reaches a state of completion, and a review paper systematically cites all the crucial papers, it becomes the representative citation for that field, eclipsing earlier papers~\cite{mcmahan2021creative}. In this way, review papers have the capacity to \emph{replace} numerous individual papers. Therefore, we can expect that while many review papers will leave the signature of developing works, some, may us say ``landmark'', review papers may exhibit the characteristics of highly disruptive papers obscuring citations toward earlier papers and making them obsolete as references.

In the WOS dataset, we selected the paper with the document type labeled as ``Review," resulting in 1,166,514 review articles. In APS dataset, review papers are selected if they are published in the journal ``Review of Modern Physics'', which is the review journal published by APS, resulting in 2,014 review articles. The results of the application of two different disruption indexes to review articles in these datasets are shown in Supplementary Figure~\ref{fig:sup_distribution_disruption_review}. In the WOS dataset, the $\Delta$ and $D$ index show different patterns. The $\Delta$ index of WOS review articles is rightward-skewed, indicating that these papers predominantly contribute to the development of their respective fields rather than being disruptive, even compared to the random baseline. On the other hand, the $D$ index of review articles exhibits a slight left skew and is also bigger than the $D$ index in the random network. In other words, $D$ index estimates the review articles as much more ``disruptive" than the baseline. In the APS dataset, both the $\Delta$ and $D$ indices reveal consistent findings: both indicate that review articles in the ``Review of Modern Physics'' have a disruptive influence on their fields, even when compared to the random baseline. 

The difference in the disruption scores of review articles in different datasets shows the complex role of review articles. The summarization of the previous works and guiding the direction of study to future generations is a unique feature that only the review article has. This can increase the awareness of prior works, but at the same time, it can also replace the reference of the previous works, which makes the future works cite the previous works less. Furthermore, the review articles, especially when authored by prominent scientists in the field, can provide a new direction of study for future works, which can change the existing flow of the evolution process of science. 

The extent of these transformative effects can be tied to the impact of the review articles themselves, particularly evident in the case of the Review of Modern Physics in the APS dataset. The papers in this journal are mostly invited and feature contributions from leading established scientists in their related fields. As reflected in its exceptionally high impact factor, exceeding 44, the papers published in this journal are significantly influential in Physics. This notable impact shows the journal’s unique position as a platform for leading scientists to share their insights and contributions to related fields. In the context of the discussion on the multifaceted role of review articles, the high disruption index of papers in Review of Modern Physics exemplifies the extent to which this distinct feature of review articles can guide and shape scientific discourse.

\subsection*{4. Additional Examples of Independent 
Simultaneous Disruption}
\label{suppinfo:additional_disruption_simultaneous}

\normalsize
\textbf{Asymptotic Freedom in Quantum Chromodynamics}

Asymptotic freedom was independently discovered by David Gross \& Frank Wilczek, and David Politzer in 1973~\cite{gross1973ultraviolet, politzer1973reliable}. Their findings demonstrated that at very short distances or high energies, the strong force weakens, enabling quarks and gluons to exhibit behavior akin to free particles, which is opposite from the earlier theory that the strong force intensifies infinitely at shorter distances. Again, they published their independent discoveries in the same issue of PRL while citing each other. This makes their disruption score low as explained earlier. The $D$ score of the paper by David J. Gross and Frank Wilczek decreased to $-0.12$ (bottom 0.4\%) from 0.05 (top 5\%) because of the citation links between them. Similarly, the $D$ score of David Politzer's paper decreased to $-0.18$ (bottom 0.2\%) from 0.15 (top 3\%). On the other hand, $\Delta$ scores for both papers are 0.98, ranking in the top 6.1\% and 5.6\%, respectively.

\textbf{Discovery of Reverse Transcriptase}

Reverse transcriptase was concurrently identified by Howard Temin and by David Baltimore, marking a pivotal moment in molecular biology in 1970~\cite{temin1970rna,baltimore1970viral}. Temin and Baltimore, working independently, published their groundbreaking findings on this unique enzyme in the journal issue, each acknowledging the other's work through mutual citations. Because of these mutual citations hiding the disruptiveness of these concurrent innovation, the disruption score of Baltimore's paper becomes $-0.45$ (bottom 0.0\%) from 0.43 (top 0.76\%) and Temin's paper becomes -0.43 (bottom 0.0\%) from 0.49 (top 0.61\%). Nevertheless, the significance of their discovery is underscored by the embedding disruptiveness measure with both papers achieving a score of 0.98 ( top 5\%).

\textbf{Higgs Mechanism}

In addition to Peter Higgs, and the group of François Englert and Robert Brout, who were mentioned in the main text, Gerald Guralnik, C. Richard Hagen, and Tom Kibble (GHK) also independently discovered the Higgs mechanism around the same time as the other teams ~\cite{englert1964broken, higgs1964broken, guralnik1964global, guralnik2009history}. However, unlike Englert and Higgs, Guralnik, Hagen, and Kibble did not receive the Nobel Prize, due to the Nobel Committee's rule of awarding the prize to no more than three individuals. As a result, their work was not included in the Nobel Prize dataset analyzed in the main text, a decision that sparked controversy within the scientific community~\cite{merali2010physicists}. The GHK group conducted their research independently, without knowledge of the other two groups' findings, and published their paper shortly after the others. Like Higgs, the GHK paper cited the simultaneous discoveries of the other teams, which contributed to a notable reduction in their $D$ score. Specifically, their paper received a $D$ score of $-0.347$, placing it in the bottom 0.04\%. Nonetheless, the $\Delta$ score for the GHK paper highlights its significant influence on the field, with a value of 0.986, ranking it in the top 5.61\%.

\subsection*{5. Additional Examples of Collective Simultaneous Disruption}
\label{suppinfo:example_collective}

\textbf{Inelastic Scattering of Electrons and Neutrons}

The deep inelastic scattering experiment conducted jointly by teams from the Massachusetts Institute of Technology (MIT) and the Stanford Linear Accelerator Center (SLAC) revealed the internal structure of protons~\cite{bloom1969high,breidenbach1969observed}. The two collaborating teams published two papers on the same experiment, citing each other. Consequently, both papers exhibit low disruption scores, measuring -0.1 and -0.06, positioning them in the bottom 0.5\% and 1.4\%, respectively. In contrast, their EDM scores stand at 0.99 and 0.98, ranking them in the top 6\% and 5.2\%, respectively.

\normalsize
\textbf{Density Functional Theory}

In 1964, Walter Kohn and Pierre Hohenberg introduced a revolutionary idea that would become the foundation of density functional theory (DFT)~\cite{hohenberg1964inhomogeneous}. They demonstrated that the electron density, a much simpler quantity to handle compared to the complex many-body wave function, is enough to uniquely determine the total energy and all other properties of a system. This breakthrough drastically simplified the study of electronic structures in materials. However, it was Kohn’s 1965 collaboration with Lu Sham that offered a practical computational method by introducing equations for noninteracting electrons, thus making DFT feasible for real-world applications~\cite{PhysRevFocus1998, kohn1965self}. Because of the successive publications, publication in 1965 citing the previous one in 1964, the disruption score of Kohn and Sham’s 1965 paper decreased from 0.96 (top 99.99\%) to -0.22 (bottom 0.18\%).

\normalsize
\textbf{The Electroweak Unification Theory}

The electroweak theory, often referred to as the Glashow–Weinberg–Salam theory, was developed through the independent efforts of Sheldon Glashow, Abdus Salam, and Steven Weinberg during the 1960s, culminating in their shared Nobel Prize in 1979~\cite{Nobel1979}. Glashow first proposed a foundational model in 1961 that laid the groundwork for unifying the weak and electromagnetic forces~\cite{glashow1961partial}. Independently, Salam and John Ward developed a similar idea in 1964, unaware of Glashow’s earlier work~\cite{Salam1964}. In 1967, Weinberg made a significant breakthrough by introducing a concrete explanation for how particles acquire mass through the Higgs mechanism followed by Salam’s independent work in 1968~\cite{weinberg1967model, Salam1968}. Although these discoveries emerged at different times, the physicists gradually became aware of each other's contributions, and their independent developments ultimately converged into a unified electroweak theory. Despite their overlapping contributions, Weinberg's 1967 paper has a disruption index of -0.06, which is the bottom 1.3\% because it cited Glashow's 1961 paper~\cite{glashow1961partial}. Without this single citation relationship, the disruption index of Weinberg's paper can be 0.57, which is top 0.44\%.

\subsection*{6. Identification of Disruptive Patents}
\label{suppinfo:patent}

\begin{figure}
    \centering
    \includegraphics[width=0.7\textwidth]{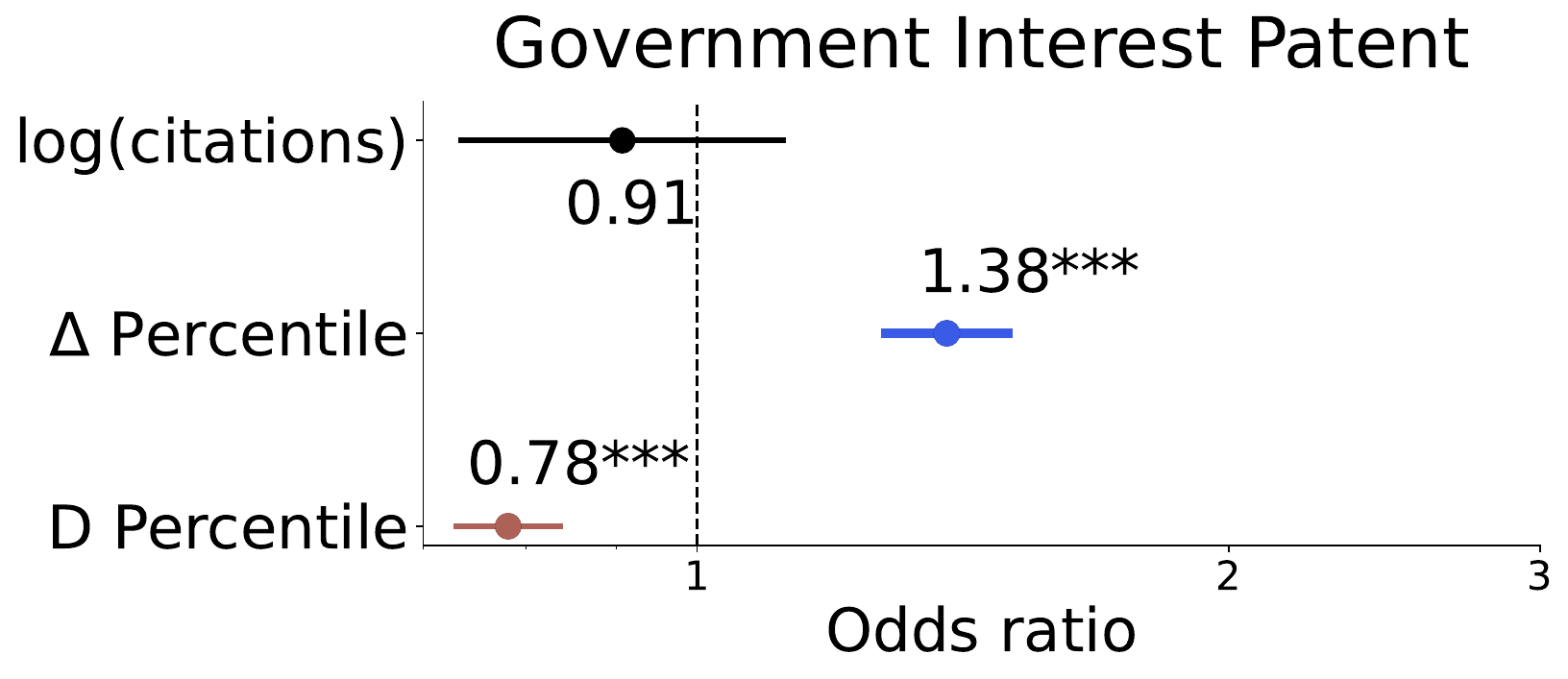}
    \caption{\textbf{The Embedding Disruptiveness Measure (EDM) Better Captures Disruptive Patents.} Firth's logistic regression on 3,018,390 patents demonstrates the higher correlation between $\Delta$ and government interest patent than the number of citations or $D$. Error bars represent the 95\% confidence interval. } 
    \label{fig:patent}
\end{figure}

We evaluated the performance of the $\Delta$ and $D$ indices in identifying disruptive work within the patent dataset by categorizing government-funded patents with a ``government interest'' acknowledgment as disruptive, building on the idea that public sector research often focuses on foundational innovations, while private companies tend to prioritize short-term, application-driven developments~\cite{partha1994toward,li2017applied,fleming2019government}. For this analysis, we used the PatentsView dataset~\cite{uspto_patentsview}, which consists of 7,387,608 patents published between 1976 and 2020. Similar to the papers dataset, we calculated the disruption index for all patents but included only those with at least 5 citations, at least 1 reference, and published between 1978 and 2018 with a utility type in the analysis, resulting in 3,018,390 patents. To assess the relationship between these indices and 56,708 government-funded patents, we performed multivariate logistic regressions using Firth's method, scaling the percentiles by a factor of 10, which is the same approach used in Section~\ref{sec:result_identification}. The independent variables in this analysis included the $D$ index percentile, $\Delta$ index percentile, and citation counts.

As a result, the odds ratio for the $\Delta$ percentile was 1.38 ($p<0.001$; 95\% CI: 1.27–1.51), indicating that a 10\% increase in the $\Delta$ percentile corresponds to a 1.38-fold increase in the odds of identifying government-funded patents. The 95\% confidence interval for this odds ratio exceeds 1, demonstrating a significant correlation between the $\Delta$ index and government interest patents. In contrast, the odds ratios for citation counts and the $D$ index percentile were 0.91 ($p>0.05$; 95\% CI: 0.73–1.12) and 0.78 ($p<0.001$; 95\% CI: 0.73–0.84), respectively, suggesting they do not have a statistically significant positive association with identifying government-funded patents. In fact, the $D$ index shows a significant negative correlation with being a government-funded patent.  

\begin{figure}[h]
    \centering
    \includegraphics[width=1\textwidth]{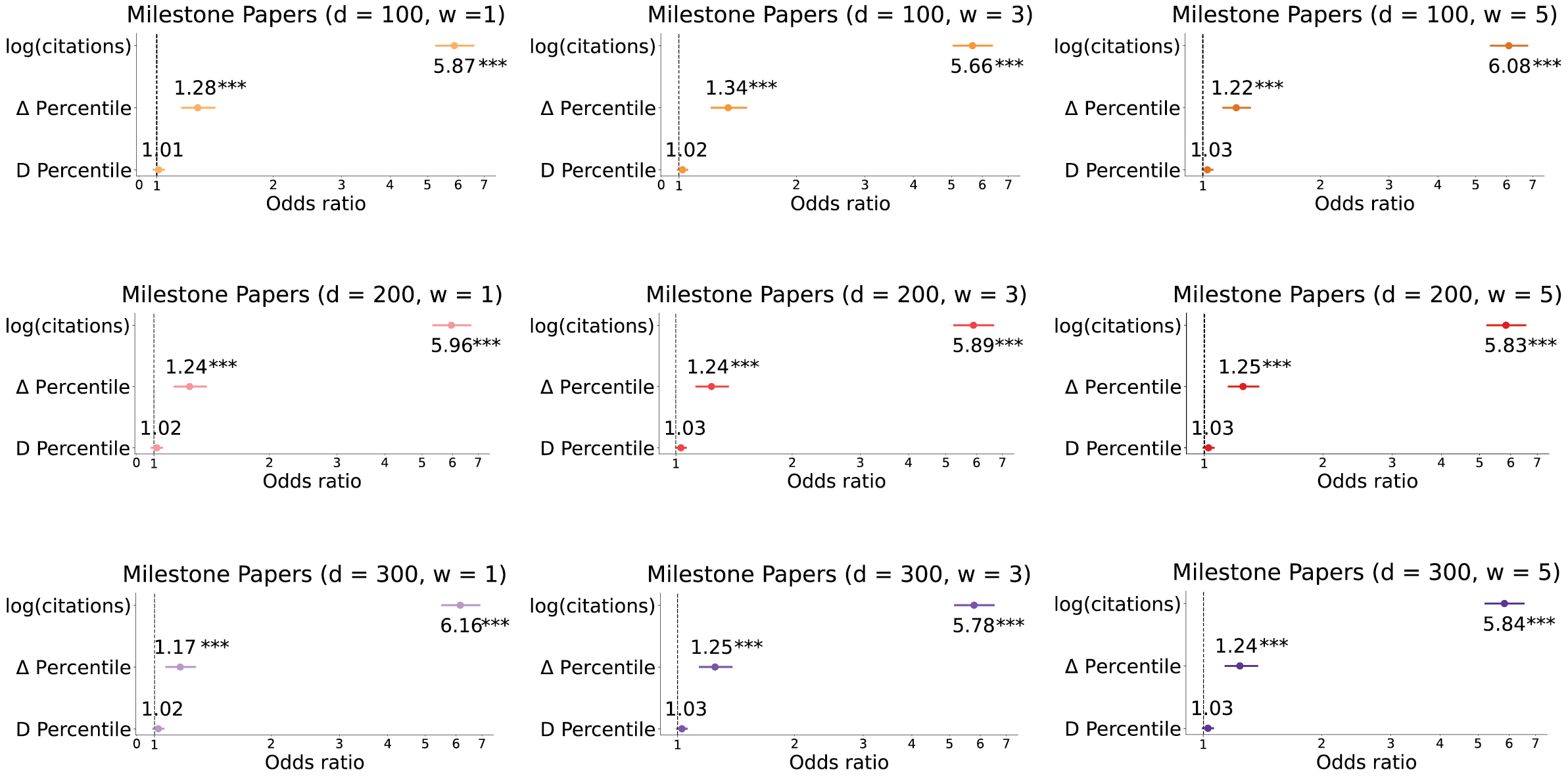}
    \caption{
\textbf{The EDM exhibits a stronger association with milestone papers than the Disruption index across various configurations of hyperparameters.} $w$ indicates the window size, which is the number of citation steps considered in the training process, and $d$ indicates the size of the embedding vector. Error bars represent the 95\% confidence interval of the odd ratios. The odd ratios of $\Delta$ calculated all variants of models are significantly bigger than 1, indicating there is a greater probability of the paper with a high $\Delta$ becoming a milestone paper. On the other hand, the odds ratios of the $D$ index are not significantly bigger than 1 across all variants. There are 278 milestone papers among 327,021 papers.}
    \label{fig:sup_robustness}
\end{figure}

\subsection*{7. Robustness across different hyper-parameters}

\normalsize 
To assess the robustness of our findings across diverse hyperparameter configurations, we conduct regression analyses on the APS dataset using various dimension and window size parameters for training the EDM model. The result is shown in Figure~\ref{fig:sup_robustness}. The $\Delta$ score for EDM exhibits superior predictive ability for milestone papers than the $D$ index consistently across different window sizes and dimensions.

\subsection*{8. Approximation of Past Vector}

Now let's consider the past vectors of $i$, $\mathbf{p}_i$. The objective function to maximize is $$\mathcal{L}_i = \sum_{j \in A_c(i)} \kappa_i^{\text{in}} \left( \mathbf{f}_j \cdot \mathbf{p}_i - \log Z_i \right) = \sum_{j \in A_c(i)} \left( \kappa_i^{\text{in}} \mathbf{f}_j \cdot \mathbf{p}_i \right) - \kappa_i^{\text{in}} |A_c(i)| \log Z_i ,$$ where $Z_i = \sum_k \exp(\mathbf{f}_k \cdot \mathbf{p}_i).$ This case is different from the future vectors because every term in $Z_i$ contains the past vector. Still, we can express $Z_i$ as following: $$Z_i = \sum_{k \in A_c(i)} \exp(\mathbf{f}_k \cdot \mathbf{p}_i) + \sum_{k' \notin A_c(i)} \exp(\mathbf{f}_{k'} \cdot \mathbf{p}_i).$$ We can ignore the constant term $\kappa_i^{\text{in}}$, and let $k \in A_c(i)$ and $k' \notin A_c(i)$. And let $M = | A_c(i) |$.  
$$\mathcal{L}_i \propto \sum_k \mathbf{f}_k \cdot \mathbf{p}_i - M \log \left(\sum_k \exp(\mathbf{f}_k \cdot \mathbf{p}_i) + \sum_{k'} \exp(\mathbf{f}_{k'} \cdot \mathbf{p}_i) \right).$$ Let's assume that the future vectors of $A_c(i)$ are closely distributed around the mean future vector $\mathbf{v}_i$. Then we can write $\mathbf{f}_k = \mathbf{v}_i + \mathbf{\epsilon}_{ik}$ for $k \in A_c(i)$. For $k' \notin A_c(i)$, we write $\mathbf{f}_{k'} = -\alpha \mathbf{v}_i + \tilde{\mathbf{f}}_{ik'}$, assuming that, if we consider all future vectors, they are isotropically distributed in the space without any preferred direction, where $\alpha \simeq \frac{1}{N-M} > 0$ and $\tilde{\mathbf{f}}_{ik'}$ are distributed isotropically. 
Then we can rewrite 

$$
\begin{aligned}\mathcal{L}_i &\propto \sum_k \mathbf{v}_i \cdot \mathbf{p}_i + \sum_k \mathbf{\epsilon}_{ik} \cdot \mathbf{p}_i - M \log \left( \sum_k \exp(\mathbf{v}_i \cdot \mathbf{p}_i + \mathbf{\epsilon}_{ik} \cdot \mathbf{p}_i) + \sum_{k'} \exp(-\alpha \mathbf{v}_i \cdot \mathbf{p}_i + \tilde{\mathbf{f}}_{ik'} \cdot \mathbf{p}_i ) \right)\\
&\simeq M \mathbf{v}_i \cdot \mathbf{p}_i - M \log S_k,
\end{aligned}$$

where 

$$\begin{aligned}S_k &= \exp(\mathbf{v}_i \cdot \mathbf{p}_i) \sum_k \exp(\mathbf{\epsilon}_{ik} \cdot \mathbf{p}_i) + \exp(-\alpha \mathbf{v}_i \cdot \mathbf{p}_i)\sum_{k'} \exp (\tilde{\mathbf{f}}_{ik'} \cdot \mathbf{p}_i)\\
&\simeq \tilde{M} \exp(\mathbf{v}_i \cdot \mathbf{p}_i) + C \exp(-\alpha \mathbf{v}_i \cdot \mathbf{p}_i),
\end{aligned}$$ 

where $\sum_k \exp(\mathbf{\epsilon}_{ik} \cdot \mathbf{p}_i) \simeq cM = \tilde{M}$ (with a constant $c$), $C = \sum_{k'} \exp(\tilde{\mathbf{f}}_{ik'} \cdot \mathbf{p}_i)$. $C$ is a function of $\Vert \mathbf{p}_i \Vert$ and can be considered as a constant given $\Vert \mathbf{p}_i \Vert$. We assume that, due to the regularizing property of stochastic optimization (both negative sampling and stochastic gradient descent), as well as the inherent noise in the data, the norm of every vector is bounded and cannot keep growing.  Now the whole function can be written as $$\mathcal{L}_i \propto s - \log (\tilde{M} e^s + C e^{-\alpha s}).$$ If we take a derivative with $s$,

$$\begin{aligned}
\frac{d \mathcal{L}_i}{d s} &= 1 - \frac{\tilde{M} e^s - \alpha C e^{-\alpha s}}{\tilde{M} e^s + C e^{-\alpha s} }\\
&=1 - \frac{(\tilde{M} e^{(1+\alpha)s} + C) - (C + \alpha C)}{\tilde{M} e^{(1+\alpha)s} + C} \\
&=\frac{(1+\alpha)C}{\tilde{M} e^{(1 + \alpha)s} + C}
.\end{aligned}$$

Given that all constants are positive, for all real values of $s$, $$\frac{d \mathcal{L}_i}{d s} > 0.$$ Therefore, $\mathcal{L}_i$ is a monotonically increasing function of $s$. In other words, given a norm $\Vert \mathbf{p}_i \Vert$, $\mathbf{p}_i$ should be as closely aligned with $\mathbf{v}_i$ as possible to maximize $\mathcal{L}_i$.

\subsection*{9. Interpretation of $\Delta$ Accounting for the Bias in Negative Sampling}

In the Methods section of the main text, we showed that the cosine distance between the future vector $\mathbf{f}_i$ and the past vector $\mathbf{p}_i$ for paper $i$ reflects the lack of reliance of its future works on its past works. However, this conclusion assumes that the embedding model is based on an unbiased word2vec approach. In reality, our method utilizes negative sampling, which introduces a bias in the probability of word2vec, as given by:

\[
\text{Pr}(v|u) = \frac{ p_0(v) \exp(\mathbf{f}_v \cdot \mathbf{p}_u)}{Z_u},
\]
where $p_0(v)$ is the noise distribution in negative sampling. Consequently, Equation (\ref{myeq1}) becomes:

\begin{align*}
  \sum_{j \in A_c(i)} \kappa_i^{\text{in}} \log \text{Pr}(j | i) + 
  \sum_{k \in D_c(i)} \kappa_{k}^{\text{in}} \log \text{Pr}(i | k)  
  &= \sum_{j \in A_c(i)} \kappa_i^{\text{in}} \left( \log p_0(i) + \log \frac{\exp({\mathbf{f}_j \cdot \mathbf{p}_i})}{Z_i} \right) \\
  &\quad + \sum_{k \in D_c(i)} \kappa_{k}^{\text{in}} \left( \log p_0(k) + \log \frac{\exp({\mathbf{f}_i \cdot \mathbf{p}_k})}{Z_k} \right) \\
  &= \sum_{j \in A_c(i)} \kappa_i^{\text{in}} \left( \log p_0(i) + \mathbf{f}_j \cdot \mathbf{p}_i - \log Z_i \right) \\
  &\quad + \sum_{k \in D_c(i)} \kappa_{k}^{\text{in}} \left( \log p_0(k) + \mathbf{f}_i \cdot \mathbf{p}_k - \log Z_k \right) \\
  & \stepcounter{equation} \tag{\theequation} \label{supp_eq_8}
\end{align*}

Since $\log p_0(i)$ and $\log p_0(k)$ are independent of $\mathbf{f}_i$ and $\mathbf{p}_i$, the derivative of Equation (\ref{supp_eq_8}) with respect to $\mathbf{f}_i$ and  $\mathbf{p}_i$ is identical to that of Equation (\ref{myeq1}). Therefore, the approximations for $\mathbf{p}_i$ and $\mathbf{f}_i$ that maximize Equation (\ref{supp_eq_8}) remain unchanged from that maximize Equation (\ref{myeq1}), and thus the resulting outcome is the same.

\subsection*{10. Proximity of Past Vectors Between Simultaneous Discovery Pairs}

In the main results, we demonstrated that the proximity of future vectors effectively identifies simultaneous discovery pairs. Here, we extend this analysis to past vectors, examining whether the simultaneous discovery pairs identified through the closest proximity of their future vectors also exhibit the closest proximity between their past vectors. Unlike the analysis of future vectors, where papers without references were included, we exclude such papers here, as past vectors are trained based on their references. The paper set used in this analysis is the same as the one used for the $\Delta$ and $D$ analysis in the Results section, consisting of 327,021 APS papers with at least five citations and one reference.

As a result, only 35.9\% of simultaneous discovery pairs have the closest proximity in their past vectors. This finding suggests that the way a paper is used by subsequent research may better reflect its unique contribution than the specific knowledge sources it initially draws upon. One possible explanation of this could be due to rhetorical citations, which are citations that serve purposes other than directly influencing new knowledge, such as providing background or context ~\cite{bao2024simulation,teplitskiy2022status}. Even if two papers of simultaneous discovery are based on the same set of foundational prior works, independent authors may use a rhetorical citation in diverse ways, meaning the set of references may differ between simultaneous discovery pairs. However, for future papers, even when a paper is cited rhetorically by future works, it is less likely that those future works would cite only one paper from a simultaneous pair without acknowledging the other, making the proximity of the future works effective to detect the simultaneous pairs.

\begin{figure}
    \centering
    \includegraphics[width=0.9\linewidth]{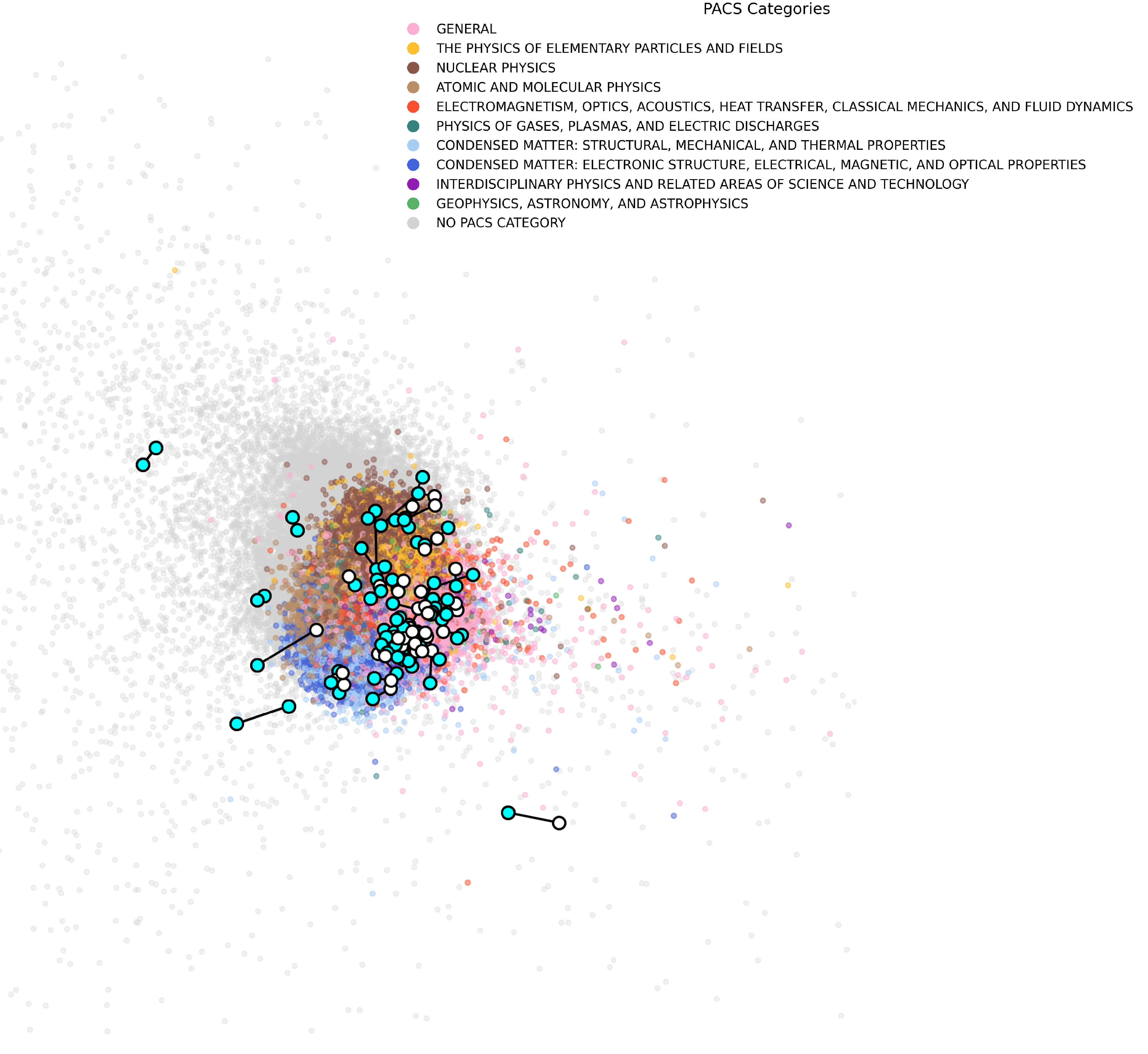}
    \caption{ \textbf{The PCA plot of the past vectors of 327,021 APS papers.} The PCA projection of each paper's past vector is color-coded based on its PACS category, with the past vectors of the 34 identified simultaneous discovery papers highlighted in bright teal. A black edge connects each paper in the simultaneous discovery pairs to its closest neighbor. If a simultaneous discovery paper's closest neighbor is not its counterpart but another paper, the closest paper is represented as a large white dot instead of teal. The plot displays principal components 2 and 3, as principal component 1 predominantly captures temporal information. }
    \label{fig:enter-label}
\end{figure}

\newpage

\renewcommand{\tablename}{Supplementary Table }

\subsection*{Supplementary Tables}
\begin{table}
    \centering
    \caption{ \textbf{Top eleven Nobel Prize-winning papers exhibiting the biggest discrepancies between their $\Delta$ and $D$ quantile rankings.} These papers are all associated with simultaneous discoveries. Higgs' paper cited Englert's paper which made disruption together~\cite{englert1964broken}, while Englert's paper did not. This makes Englert's paper maintain a high $D$ score and, as a result, Englert's paper is not included in the table. } \label{tab:d_and_delta_nobel}
    \footnotesize
    \begin{tabularx}{\linewidth}{|p{7 cm}|c|c|p{4 cm}} 
        \toprule
        \textbf{Paper} &  \(\mathbf{D}\) quantile & \(\mathbf{\Delta}\) quantile
        & \textbf{Simultaneous Discovery Pair}
        \\
        \midrule
   \footnotesize
 Kohn, W., \& Sham, L. J. (1965). Self-consistent equations including  exchange and correlation effects. Physical review, 140(4A), A1133.& 0.001&0.959 & 
Hohenberg, P., \& Kohn, W. (1964). \\ 
    \hline

  Higgs, P. W. (1964). Broken symmetries and the masses of gauge bosons. Physical review letters, 13(16), 508. & 0.001& 0.959 & Englert, F., \& Brout, R. (1964). Guralnik, G. S., Hagen, C. R., \& Kibble, T. W. B. (1964). \\ 
    \hline
    Augustin, J. E., et al. (1974). Discovery of a Narrow Resonance in e+ e- Annihilation.   Physical Review Letters, 33(23), 1406.& 0.000&  0.952&Abrams, G. S., et al. (1974).\\ 
    \hline
      Baltimore, D. (1970). Viral RNA-dependent DNA polymerase: RNA-dependent DNA polymerase in virions of RNA tumour viruses. Nature, 226(5252), 1209-1211. &0.000 &0.947 & Temin, H. M., \& Mizutami, S. (1970) \\ 
    \hline
     Temin, H. M., \& Mizutami, S. (1970). RNA-dependent DNA polymerase in virions of Rous sarcoma virus. Nature, 226, 1211-1213. &0.000 &0.945
 & Baltimore, D. (1970). \\ 
    \hline
 Weinberg, S. (1967). A model of leptons. Physical review letters, 19(21), 1264. &0.013 &0.956& Sheldon L. Glashow (1961).  \\ 
    \hline
       
       Gross, D. J., \& Wilczek, F. (1973). Ultraviolet behavior of non-abelian gauge theories. Physical Review Letters, 30(26), 1343. & 0.004& 0.946& Politzer, H. D. (1973)\\ 
    \hline
     
    Politzer, H. D. (1973). Reliable perturbative results for strong interactions? Physical Review Letters, 30(26), 1346. &0.002 &  0.942 &Gross, D. J., \& Wilczek, F. (1973)\\ 
       \hline
        Bloom, E. D., et al. (1969). High-Energy Inelastic e- p Scattering at 6 and 10. Physical Review Letters, 23(16), 930. & 0.013 & 0.951&Breidenbach, M., et al. (1969).\\ 
    
    \hline
        Breidenbach, M., et al. (1969). Observed behavior of highly inelastic electron-proton scattering. Physical Review Letters, 23(16), 935. & 0.005 &0.942 & Bloom, E. D., et al. (1969). \\ 
    \hline
    Abrams, G. S., et al. (1974). Discovery of a second narrow resonance in e+ e- annihilation. Physical Review Letters, 33(24), 1453. &0.001 & 0.936& Augustin, J. E., et al. (1974).\\ 
    \hline
\bottomrule
    \end{tabularx}
\end{table}

\newpage

\printbibliography[heading=subbibliography, title={Supplementary Information References}]

@article{glashow1961partial,
  title={Partial-symmetries of weak interactions},
  author={Glashow, Sheldon L},
 journal={Nucl. Phys.},
  volume={22},
  number={4},
  pages={579--588},
  year={1961},
  publisher={Elsevier}
}

@article{fleming2019government,
author = {L. Fleming  and H. Greene  and G. Li  and M. Marx  and D. Yao },
title = {Government-funded research increasingly fuels innovation},
journal = {Science},
volume = {364},
number = {6446},
pages = {1139-1141},
year = {2019},
doi = {10.1126/science.aaw2373},
URL = {https://www.science.org/doi/abs/10.1126/science.aaw2373},
eprint = {https://www.science.org/doi/pdf/10.1126/science.aaw2373}}

@article{li2017applied,
  author = {Danielle Li  and Pierre Azoulay  and Bhaven N. Sampat },
title = {The applied value of public investments in biomedical research},
journal = {Science},
volume = {356},
number = {6333},
pages = {78-81},
year = {2017},
doi = {10.1126/science.aal0010},
URL = {https://www.science.org/doi/abs/10.1126/science.aal0010},
eprint = {https://www.science.org/doi/pdf/10.1126/science.aal0010}}

@article{partha1994toward,
  title={Toward a new economics of science},
  author={Partha, Dasgupta and David, Paul A},
  journal={Res. Policy},
  volume={23},
  number={5},
  pages={487--521},
  year={1994},
  publisher={Elsevier},
doi = {10.1016/0048-7333(94)01002-1}
}

@article{hohenberg1964inhomogeneous,
  title={Inhomogeneous electron gas},
  author={Hohenberg, Pierre and Kohn, Walter},
  journal={Phys. Rev},
  volume={136},
  number={3B},
  pages={B864},
  year={1964},
  publisher={APS},
doi ={10.1103/PhysRev.136.B864}
}

@misc{Nobel1979,
  title        = {The Nobel Prize in Physics 1979},
  author       = {{Nobel Prize Outreach AB}},
  year         = {2024},
  note         = { Accessed: 6 Sep 2024},
  url          = {https://www.nobelprize.org/prizes/physics/1979/summary/},
 howpublished = {\url{https://www.nobelprize.org/prizes/physics/1979/summary/}},
}

@article{bao2024simulation,
  title={A simulation-based analysis of the impact of rhetorical citations in science},
  author={Bao, Honglin and Teplitskiy, Misha},
 journal={Nat. Commun.},
  volume={15},
  number={1},
  pages={431},
  year={2024},
  publisher={Nature Publishing Group UK London},
doi = {10.1038/s41467-023-44249-0}
}

@article{teplitskiy2022status,
  title={How status of research papers affects the way they are read and cited},
  author={Teplitskiy, Misha and Duede, Eamon and Menietti, Michael and Lakhani, Karim R},
 journal={Res. Policy},
  volume={51},
  number={4},
  pages={104484},
  year={2022},
  publisher={Elsevier},
doi = {10.1016/j.respol.2022.104484}

}

@article{PhysRevFocus1998,
  title     = {Nobel Focus: Chemistry by Computer},
  author    = {American Physical Society},
  journal   = {Phys. Rev. Focus},
  volume    = {2},
  pages     = {19},
  year      = {1998},
  month     = {10},
  note      = {Accessed: 05 Sep 2024},
  url       = {https://physics.aps.org/story/v2/st19}
}

@inproceedings{Salam1968,
  author    = {A. Salam},
  title     = {In Elementary Particle Physics (Nobel Symposium No. 8)},
  booktitle = {Almqvist and Wilsell},
  year      = {1968},
  address   = {Stockholm}
}

@article{Salam1964,
  author    = {A. Salam and J.C. Ward},
  title     = {Electromagnetic and weak interactions},
  journal   = {Phys. Lett.},
  year      = {1964},
  volume    = {13},
  number    = {2},
  pages     = {168--171},
  doi       = {10.1016/0031-9163(64)90711-5}
}

@article{painter2020quantifying,
 title = {Quantifying simultaneous innovations in evolutionary medicine},
	volume = {139},
	issn = {1611-7530},
	url = {https://doi.org/10.1007/s12064-020-00333-3},
	doi = {10.1007/s12064-020-00333-3},
	number = {4},
	journal = {Theory Biosci.},
	author = {Painter, Deryc T. and van der Wouden, Frank and Laubichler, Manfred D. and Youn, Hyejin},
	month = {12},
	year = {2020},
	pages = {319--335},
}

@article{weinberg1967model,
  title={A model of leptons},
  author={Weinberg, Steven},
  journal={Phys. Rev. Lett.},
  volume={19},
  number={21},
  pages={1264},
  year={1967},
  publisher={APS},
 doi ={10.1103/PhysRevLett.19.1264}
}

@Manual{ firth_package,
  title = {Package `logistf': Firth's Bias-Reduced Logistic Regression},
  author = {Georg Heinze and Meinhard Ploner},
  year = {2023},
  note = {Version 1.26.0},
  month = {8},
  url = {https://cran.r-project.org/package=logistf},
  description = {Fit a logistic regression model using Firth's bias reduction method, equivalent to penalization of the log-likelihood by the Jeffreys prior. Confidence intervals for regression coefficients can be computed by penalized profile likelihood. Firth's method was proposed as an ideal solution to the problem of separation in logistic regression, see Heinze and Schemper (2002). If needed, the bias reduction can be turned off to obtain ordinary maximum likelihood logistic regression.},
  depends = {R (>= 3.0.0)},
  imports = {mice, mgcv, formula.tools, Matrix},
  suggests = {emmeans (>= 1.4), estimability}
}

@article{firth1993bias,
  title={Bias reduction of maximum likelihood estimates},
 ISSN = {00063444},
 URL = {http://www.jstor.org/stable/2336755},
 author = {David Firth},
 journal = {Biometrika},
 number = {1},
 pages = {27--38},
 publisher = {[Oxford University Press, Biometrika Trust]},
 title = {Bias Reduction of Maximum Likelihood Estimates},
 urldate = {2024-11-20},
 volume = {80},
 year = {1993},
doi = {10.2307/2336755}
}

@article{qu2024outliers,
author = {Qu, Guannan and Chen, Kaihua and Wang, Luyao and Yang, Yayu and Zhang, Ruhao},
title = {Are outliers more disruptive? Technological niche, disruptive innovation, and recombinant capability},
journal = {R\&D Management},
pages = {},
doi = {https://doi.org/10.1111/radm.12685},
url = {https://onlinelibrary.wiley.com/doi/abs/10.1111/radm.12685}
,
year = {2024}}

@article{ruan2021rethinking,
    author={Ruan, Xuanmin and Lyu, Dongqing and Gong, Kaile and Cheng, Ying and Li, Jiang},
  title={{Rethinking the disruption index as a measure of scientific and technological advances}},
  journal={Technol. Forecast. Soc. Change
},
  year=2021,
  volume={172},
  number={C},
  pages={},
  month={},
  doi={10.1016/j.techfore.2021.1},
  url={https://ideas.repec.org/a/eee/tefoso/v172y2021ics0040162521005035.html},
doi = {10.1016/j.techfore.2021.121071}
}

@article{augustin1974discovery,
   title = {Discovery of a Narrow Resonance in ${e}^{+}{e}^{\ensuremath{-}}$ Annihilation},
  author = {Augustin, J.-E. and Boyarski, A. M. and Breidenbach, M. and Bulos, F. and Dakin, J. T. and Feldman, G. J. and Fischer, G. E. and Fryberger, D. and Hanson, G. and Jean-Marie, B. and Larsen, R. R. and L\"uth, V. and Lynch, H. L. and Lyon, D. and Morehouse, C. C. and Paterson, J. M. and Perl, M. L. and Richter, B. and Rapidis, P. and Schwitters, R. F. and Tanenbaum, W. M. and Vannucci, F. and Abrams, G. S. and Briggs, D. and Chinowsky, W. and Friedberg, C. E. and Goldhaber, G. and Hollebeek, R. J. and Kadyk, J. A. and Lulu, B. and Pierre, F. and Trilling, G. H. and Whitaker, J. S. and Wiss, J. and Zipse, J. E.},
  journal = {Phys. Rev. Lett.},
  volume = {33},
  issue = {23},
  pages = {1406--1408},
  numpages = {0},
  year = {1974},
  month = {12},
  publisher = {American Physical Society},
  doi = {10.1103/PhysRevLett.33.1406},
  url = {https://link.aps.org/doi/10.1103/PhysRevLett.33.1406}
}

@article{aubert1974experimental,
   title = {Experimental Observation of a Heavy Particle $J$},
  author = {Aubert, J. J. and Becker, U. and Biggs, P. J. and Burger, J. and Chen, M. and Everhart, G. and Goldhagen, P. and Leong, J. and McCorriston, T. and Rhoades, T. G. and Rohde, M. and Ting, Samuel C. C. and Wu, Sau Lan and Lee, Y. Y.},
  journal = {Phys. Rev. Lett.},
  volume = {33},
  issue = {23},
  pages = {1404--1406},
  numpages = {0},
  year = {1974},
  month = {12},
  publisher = {American Physical Society},
  doi = {10.1103/PhysRevLett.33.1404},
  url = {https://link.aps.org/doi/10.1103/PhysRevLett.33.1404}
}

@article{kohn1965self,
 title = {Self-Consistent Equations Including Exchange and Correlation Effects},
  author = {Kohn, W. and Sham, L. J.},
  journal = {Phys. Rev.},
  volume = {140},
  issue = {4A},
  pages = {A1133--A1138},
  numpages = {0},
  year = {1965},
  month = {11},
  publisher = {American Physical Society},
  doi = {10.1103/PhysRev.140.A1133},
  url = {https://link.aps.org/doi/10.1103/PhysRev.140.A1133}
}

@book{schumpeter2013capitalism,
  title={Capitalism, socialism and democracy},
  author={Schumpeter, Joseph A},
  year={2013},
  publisher={Routledge}
}

@book{kuhn2012structure,
  title={The structure of scientific revolutions},
  author={Kuhn, Thomas S},
  year={2012},
  publisher={University of Chicago press}
}

@misc{prl_milestone,
  title = {Letters from the Past - A PRL Retrospective},
  howpublished = {\url{https://journals.aps.org/prl/50years/milestones}},
  note = {Accessed: 30 Sep 2023}
}

@misc{pre_milestone,
  title = {Physical Review E Milestones},
  howpublished = {\url{https://journals.aps.org/pre/collections/pre-milestones}},
  note = {Accessed: 30 Sep 2023}
}

@misc{prd_milestone,
  title = {Physical Review D 50th Anniversary Milestones},
  howpublished = {\url{https://journals.aps.org/prd/50th}},
  note = {Accessed: 30 Sep 2023}
}

@misc{prc_milestone,
  title = {Physical Review C 50th Anniversary Milestones},
  howpublished = {\url{https://journals.aps.org/prc/50th}},
  note = {Accessed: 30 Sep 2023}
}

@misc{prb_milestone,
  title = {Physical Review B 50th Anniversary Milestones},
  howpublished = {\url{https://journals.aps.org/prb/50th}},
  note = {Accessed: 30 Sep 2023}
}

@misc{pra_milestone,
  title = {Physical Review A 50th Anniversary Milestones},
  howpublished = {\url{https://journals.aps.org/pra/50th}},
  note = {Accessed: 30 Sep 2023}
}

@misc{NSF_nobel,
  title = {The Nobel Prizes: The NSF Connection},
  howpublished = {\url{https://www.nsf.gov/news/special_reports/nobelprizes/}},
  note = {Accessed: 30 Sep 2023}
}

@article{DVN/6NJ5RN_2018,
 title = {A dataset of publication records for Nobel laureates},
	volume = {6},
	issn = {2052-4463},
	url = {https://doi.org/10.1038/s41597-019-0033-6},
	doi = {10.1038/s41597-019-0033-6},
	pages = {33},
	number = {1},
	journaltitle = {Scientific Data},
	shortjournal = {Scientific Data},
	author = {Li, Jichao and Yin, Yian and Fortunato, Santo and Wang, Dashun},
	date = {2019-04-18},
}

@article{ioannidis2020work,
  title={Work honored by Nobel prizes clusters heavily in a few scientific fields},
  author={Ioannidis, John PA and Cristea, Ioana-Alina and Boyack, Kevin W},
  journal={PLOS ONE},
  volume={15},
  number={7},
  pages={e0234612},
  year={2020},
  publisher={Public Library of Science San Francisco, CA USA},
doi = {10.1371/journal.pone.0234612}
}

@article{funk2017dynamic,
author = {Funk, Russell J. and Owen-Smith, Jason},
title = {A Dynamic Network Measure of Technological Change},
journal = {Manage. Sci.},
volume = {63},
number = {3},
pages = {791-817},
year = {2017},
doi = {10.1287/mnsc.2015.2366},
URL = { 
    
        https://doi.org/10.1287/mnsc.2015.2366


}
}

@misc{uspto_patentsview,
  author       = {{U.S. Patent and Trademark Office}},
  title        = "{Data Download Tables}",
 howpublished = {\url{https://patentsview.org/download/data-download-tables}},
  note         = "Accessed: 05 Nov 2024"
}

@article{park2023papers,
  title={Papers and patents are becoming less disruptive over time},
  	volume = {613},
	issn = {1476-4687},
	url = {https://doi.org/10.1038/s41586-022-05543-x},
	doi = {10.1038/s41586-022-05543-x},
	number = {7942},
	journal = {Nature},
	author = {Park, Michael and Leahey, Erin and Funk, Russell J.},
	month = jan,
	year = {2023},
	pages = {138--144},
}

@article{mcmahan2021creative,
author = {Peter McMahan and Daniel A. McFarland},
title ={Creative Destruction: The Structural Consequences of Scientific Curation},

journal = {	Am. Sociol. Rev.},
volume = {86},
number = {2},
pages = {341-376},
year = {2021},
doi = {10.1177/0003122421996323}
}

@article{lin2022new,
title = {New directions in science emerge from disconnection and discord},
journal = {	J. Informetr.},
volume = {16},
number = {1},
pages = {101234},
year = {2022},
issn = {1751-1577},
doi = {https://doi.org/10.1016/j.joi.2021.101234},
url = {https://www.sciencedirect.com/science/article/pii/S175115772100105X},
author = {Yiling Lin and James A. Evans and Lingfei Wu}
}

@article{petersen2023disruption,
  	title = {The disruption index is biased by citation inflation},
	issn = {2641-3337},
	url = {https://doi.org/10.1162/qss\_a\_00333},
	doi = {10.1162/qss_a_00333},
	journal = {Quant. Sci. Stud.},
	author = {Petersen, Alexander Michael and Arroyave, Felber and Pammolli, Fabio},
	month = {11},
	year = {2024},
	pages = {1--18},
}

@article{guralnik2009history,
  title={The history of the Guralnik, Hagen and Kibble development of the theory of spontaneous symmetry breaking and gauge particles},
  author={Guralnik, Gerald S},
  journal={	Int. J. Mod. Phys. A},
  volume={24},
  number={14},
  pages={2601--2627},
  year={2009},
  publisher={World Scientific},
doi = {10.1142/S0217751X09045431},
}

@article{mikolov2013distributed,
  title={Distributed representations of words and phrases and their compositionality},
  author={Mikolov, Tomas and Sutskever, Ilya and Chen, Kai and Corrado, Greg S and Dean, Jeff},
  journal={Advances in Neural Information Processing Systems},
  volume={26},
  year={2013}
}

@article{gross1973ultraviolet,
  title={Ultraviolet behavior of non-abelian gauge theories},
  author={Gross, David J and Wilczek, Frank},
  journal={Phys. Rev. Lett.},
  volume={30},
  number={26},
  pages={1343},
  year={1973},
  publisher={APS},
doi = {10.1103/PhysRevLett.30.1343}
}

@article{kojaku2021residual2vec,
  title={Residual2Vec: Debiasing graph embedding with random graphs},
  author={Kojaku, Sadamori and Yoon, Jisung and Constantino, Isabel and Ahn, Yong-Yeol},
  journal={Advances in Neural Information Processing Systems},
  volume={34},
  pages={24150--24163},
  year={2021},

}

@article{bittmann2021applied,
 	title = {Applied usage and performance of statistical matching in bibliometrics: {The} comparison of milestone and regular papers with multiple measurements of disruptiveness as an empirical example},
	volume = {2},
	issn = {2641-3337},
	url = {https://doi.org/10.1162/qss\_a\_00158},
	doi = {10.1162/qss_a_00158},
	number = {4},
	journal = {Quant. Sci. Stud.},
	author = {Bittmann, Felix and Tekles, Alexander and Bornmann, Lutz},
	month = dec,
	year = {2021},
	pages = {1246--1270},
}

@article{schakel2015measuring,
  title={Measuring word significance using distributed representations of words},
  author={Schakel, Adriaan MJ and Wilson, Benjamin J},
  note={Preprint at \url{https://arxiv.org/abs/1508.02297}},
  year={2015}
}

@article{murray2020unsupervised,
  author    = {Murray, D. and Yoon, J. and Kojaku, S. and Costas, R. and Jung, W. and Milojević, S. and Ahn, Y.},
  title     = {Unsupervised embedding of trajectories captures the latent structure of scientific migration},
  journal   = {Proc. Natl Acad. Sci. USA},
  volume    = {120},
  number    = {52},
  pages     = {e2305414120},
  year      = {2023}
}

@inproceedings{kim2022quantifying,
author = {Kim, Munjung and Yoon, Jisung and Jung, Woo-Sung and Kim, Hyunuk},
title = {Quantifying the Topic Disparity of Scientific Articles},
year = {2022},
isbn = {9781450391306},
publisher = {Association for Computing Machinery},
address = {New York, NY, USA},
url = {https://doi.org/10.1145/3487553.3524655},
doi = {10.1145/3487553.3524655},
booktitle = {Companion Proceedings of the Web Conference 2022},
pages = {769–773},
numpages = {5},
location = {Virtual Event, Lyon, France},
series = {WWW '22}
}

@article{bornmann2021convergent,
 title = {Convergent validity of several indicators measuring disruptiveness with milestone assignments to physics papers by experts},
journal = {	J. Informetr.},
volume = {15},
number = {3},
pages = {101159},
year = {2021},
issn = {1751-1577},
doi = {https://doi.org/10.1016/j.joi.2021.101159},
url = {https://www.sciencedirect.com/science/article/pii/S1751157721000304},
author = {Lutz Bornmann and Alexander Tekles}
}

@article{merali_physicists_2010,
	title = {Physicists get political over Higgs},
	url = {https://doi.org/10.1038/news.2010.390},
	doi = {10.1038/news.2010.390},
	journaltitle = {Nature},
	shortjournal = {Nature},
	author = {Merali, Zeeya},
	date = {2010-08-04},
}

@misc{APS125Years,
  title = {125 Years of the American Physical Society Journals},
  author = {{American Physical Society}},
  year = {2024},
  howpublished = {\url{https://journals.aps.org/125years}},
  note = {Accessed: 2024-11-05}
}

@article{kedrick2024conceptual,
	title = {Conceptual structure and the growth of scientific knowledge},
	volume = {8},
	issn = {2397-3374},
	url = {https://doi.org/10.1038/s41562-024-01957-x},
	doi = {10.1038/s41562-024-01957-x},
	number = {10},
	journal = {Nat. Hum. Behav.},
	author = {Kedrick, Kara and Levitskaya, Ekaterina and Funk, Russell J.},
	month = {10},
	year = {2024},
	pages = {1915--1923},
}

@article{lin2023remote,
 title = {Remote collaboration fuses fewer breakthrough ideas},
	volume = {623},
	issn = {1476-4687},
	url = {https://doi.org/10.1038/s41586-023-06767-1},
	doi = {10.1038/s41586-023-06767-1},
	number = {7989},
	journal = {Nature},
	author = {Lin, Yiling and Frey, Carl Benedikt and Wu, Lingfei},
	month = nov,
	year = {2023},
	pages = {987--991},
}

@article{milojevic2015quantifying,
  title = {Quantifying the cognitive extent of science},
journal = {	J. Informetr.},
volume = {9},
number = {4},
pages = {962-973},
year = {2015},
issn = {1751-1577},
doi = {https://doi.org/10.1016/j.joi.2015.10.005},
url = {https://www.sciencedirect.com/science/article/pii/S175115771530081X},
author = {Staša Milojević},
}

@article{fortunato2007random,
  title={Random walks on directed networks: the case of PageRank},
  author={Fortunato, Santo and Flammini, Alessandro},
  journal={Int. J. Bifurc. Chaos Appl. Sci. Eng.},
  volume={17},
  number={07},
  pages={2343--2353},
  year={2007},
  publisher={World Scientific}
}

@article{zeng2021fresh,
title = {Fresh teams are associated with original and multidisciplinary research},
	volume = {5},
	issn = {2397-3374},
	url = {https://doi.org/10.1038/s41562-021-01084-x},
	doi = {10.1038/s41562-021-01084-x},
	number = {10},
	journal = {Nat. Hum. Behav. },
	author = {Zeng, An and Fan, Ying and Di, Zengru and Wang, Yougui and Havlin, Shlomo},
	month = {10},
	year = {2021},
	pages = {1314--1322},
}

@article{paul2021writing,
  title={Writing an impactful review article: what do we know and what do we need to know?},
  author={Paul, Justin and Merchant, Altaf and Dwivedi, Yogesh K and Rose, Gregory},
  journal={	J. Bus. Res.},
  volume={133},
  pages={337--340},
  year={2021},
  publisher={Elsevier}
}

@misc{palmatier2018review,
  title={Review articles: purpose, process, and structure},
  author={Palmatier, Robert W and Houston, Mark B and Hulland, John},
  journal={J. Acad. Mark. Sci.},
  volume={46},
  pages={1--5},
  year={2018},
  publisher={Springer}
}

@article{merali2010physicists,
  author       = {Zeeya Merali},
  title        = {Physicists get political over {Higgs}},
  journal      = {Nature News},
  year         = {2010},
  month        = {8},
  day          = {4},
  url          = {https://doi.org/10.1038/news.2010.390},
  doi          = {10.1038/news.2010.390},
}

@article{masuda2017random,
  title={Random walks and diffusion on networks},
  author={Masuda, Naoki and Porter, Mason A and Lambiotte, Renaud},
  journal={Phys. Rep.},
  volume={716},
  pages={1--58},
  year={2017},
  publisher={Elsevier}
}

@article{bentley2023disruption,
  title={Is disruption decreasing, or is it accelerating?},
  author    = {Bentley, R. Alexander and Valverde, Sergi and Borycz, Joshua and Vidiella, Blai and Horne, Benjamin D. and Duran-Nebreda, Salva and O'Brien, Michael J.},
  journal   = {Adv. Complex Syst.},
  volume    = {26},
  number    = {02},
  year      = {2023}
}

@inproceedings{tang2015line,
  title={Line: Large-scale information network embedding},
  author={Tang, Jian and Qu, Meng and Wang, Mingzhe and Zhang, Ming and Yan, Jun and Mei, Qiaozhu},
  booktitle={Proceedings of the 24th international conference on world wide web},
  pages={1067--1077},
  year={2015}
}

@inproceedings{perozzi2014deepwalk,
  title={Deepwalk: Online learning of social representations},
  author={Perozzi, Bryan and Al-Rfou, Rami and Skiena, Steven},
  booktitle={Proceedings of the 20th ACM SIGKDD international conference on Knowledge discovery and data mining},
  pages={701--710},
  year={2014}
}

@article{aidelsburger2013realization,
    title = {Realization of the Hofstadter Hamiltonian with Ultracold Atoms in Optical Lattices},
  author = {Aidelsburger, M. and Atala, M. and Lohse, M. and Barreiro, J. T. and Paredes, B. and Bloch, I.},
  journal = {Phys. Rev. Lett.},
  volume = {111},
  issue = {18},
  pages = {185301},
  numpages = {5},
  year = {2013},
  month = {10},
  publisher = {American Physical Society},
  doi = {10.1103/PhysRevLett.111.185301},
  url = {https://link.aps.org/doi/10.1103/PhysRevLett.111.185301}
}

@article{miyake2013realizing,
  title = {Realizing the Harper Hamiltonian with Laser-Assisted Tunneling in Optical Lattices},
  author = {Miyake, Hirokazu and Siviloglou, Georgios A. and Kennedy, Colin J. and Burton, William Cody and Ketterle, Wolfgang},
  journal = {Phys. Rev. Lett.},
  volume = {111},
  issue = {18},
  pages = {185302},
  numpages = {5},
  year = {2013},
  month = {10},
  publisher = {American Physical Society},
  doi = {10.1103/PhysRevLett.111.185302},
  url = {https://link.aps.org/doi/10.1103/PhysRevLett.111.185302}
}

@inproceedings{grover2016node2vec,
author = {Grover, Aditya and Leskovec, Jure},
title = {node2vec: Scalable Feature Learning for Networks},
year = {2016},
publisher = {Association for Computing Machinery},
address = {New York, NY, USA},
url = {https://doi.org/10.1145/2939672.2939754},
doi = {10.1145/2939672.2939754},
booktitle = {Proceedings of the 22nd ACM SIGKDD International Conference on Knowledge Discovery and Data Mining},
pages = {855–864},
numpages = {10},
location = {San Francisco, California, USA},
series = {KDD '16}
}

@article{small1973co,
  title={Co-citation in the scientific literature: A new measure of the relationship between two documents},
  author={Small, Henry},
  journal={J. Am. Soc. Info. Sci.},
  volume={24},
  number={4},
  pages={265--269},
  year={1973},
  publisher={Wiley Online Library}
}

@article{simonton1979multiple,
  title={Multiple discovery and invention: Zeitgeist, genius, or chance?},
  author={Simonton, Dean K},
  year={1979},
  publisher={American Psychological Association},
journal= {J. Pers. Soc. Psychol.},
volume={37},
  number={9},
  pages={1603--1616},
  doi={10.1037/0022-3514.37.9.1603}
}

@article{wu2019solo,
  title={Solo citations, duet citations, and prelude citations: New measures of the disruption of academic papers},
  author={Wu, Qiang and Yan, Zhaoyang},
  note={Preprint at \url{https://arxiv.org/abs/1905.03461}},
  year={2019}
}

@article{merton1961singletons,
  title={Singletons and multiples in scientific discovery: A chapter in the sociology of science},
  author={Merton, Robert K},
  journal={Proc. Am. Philos. Soc. },
  volume={105},
  number={5},
  pages={470--486},
  year={1961},
  publisher={JSTOR}
}

@article{higgs1964broken,
 title = {Broken Symmetries and the Masses of Gauge Bosons},
  author = {Higgs, Peter W.},
  journal = {Phys. Rev. Lett.},
  volume = {13},
  issue = {16},
  pages = {508--509},
  numpages = {0},
  year = {1964},
  month = {10},
  publisher = {American Physical Society},
  doi = {10.1103/PhysRevLett.13.508},
  url = {https://link.aps.org/doi/10.1103/PhysRevLett.13.508}
}

@inproceedings{song2018directional,
    title = "Directional Skip-Gram: Explicitly Distinguishing Left and Right Context for Word Embeddings",
    author = "Song, Yan  and
      Shi, Shuming  and
      Li, Jing  and
      Zhang, Haisong",
    editor = "Walker, Marilyn  and
      Ji, Heng  and
      Stent, Amanda",
    booktitle = "Proceedings of the 2018 Conference of the North {A}merican Chapter of the Association for Computational Linguistics: Human Language Technologies, Volume 2 (Short Papers)",
    month = jun,
    year = "2018",
    address = "New Orleans, Louisiana",
    publisher = "Association for Computational Linguistics",
    url = "https://aclanthology.org/N18-2028",
    doi = "10.18653/v1/N18-2028",
    pages = "175--180",
}

@article{guralnik1964global,
  title={Global conservation laws and massless particles},
  author={Guralnik, Gerald S and Hagen, Carl R and Kibble, Thomas WB},
  journal={Phys. Rev. Lett.},
  volume={13},
  number={20},
  pages={585},
  year={1964},
  publisher={APS}
}

@article{bikard2020idea,
author = {Bikard, Michaël},
title = {Idea twins: Simultaneous discoveries as a research tool},
journal = {Strateg. Manag. J.},
volume = {41},
number = {8},
pages = {1528-1543},
doi = {https://doi.org/10.1002/smj.3162},
url = {https://onlinelibrary.wiley.com/doi/abs/10.1002/smj.3162},
eprint = {https://onlinelibrary.wiley.com/doi/pdf/10.1002/smj.3162},
year = {2020}
}

@article{breidenbach1969observed,
  title={Observed behavior of highly inelastic electron-proton scattering},
  author={Breidenbach, Martin and Friedman, Jerome I and Kendall, Henry W and Bloom, Elliott D and Coward, DH and DeStaebler, H and Drees, J and Mo, Luke W and Taylor, Richard E},
  journal={Phys. Rev. Lett.},
  volume={23},
  number={16},
  pages={935},
  year={1969},
  publisher={APS}
}

@article{bloom1969high,
  title={High-Energy Inelastic e- p Scattering at 6 and 10},
  author={Bloom, Elliott D and Coward, DH and DeStaebler, H and Drees, J and Miller, Guthrie and Mo, Luke W and Taylor, Richard E and Breidenbach, Martin and Friedman, Jerome I and Hartmann, George C and others},
  journal={Phys. Rev. Lett.},
  volume={23},
  number={16},
  pages={930},
  year={1969},
  publisher={APS}
}

@article{politzer1973reliable,
  title={Reliable perturbative results for strong interactions?},
  author={Politzer, H David},
  journal={Phys. Rev. Lett.},
  volume={30},
  number={26},
  pages={1346},
  year={1973},
  publisher={APS}
}

@article{temin1970rna,
  title={RNA-dependent DNA polymerase in virions of Rous sarcoma virus.},
  author={Temin, Howard M and Mizutami, S and others},
  journal={Nature},
  volume={226},
  pages={1211--1213},
  year={1970},
  publisher={Lond.}
}

@article{baltimore1970viral,
  title={Viral RNA-dependent DNA polymerase: RNA-dependent DNA polymerase in virions of RNA tumour viruses},
  author={Baltimore, David},
  journal={Nature},
  volume={226},
  number={5252},
  pages={1209--1211},
  year={1970},
  publisher={Nature Publishing Group UK London}
}

@article{englert1964broken,
  title={Broken symmetry and the mass of gauge vector mesons},
  author={Englert, Fran{\c{c}}ois and Brout, Robert},
  journal={Phys. Rev. Lett.},
  volume={13},
  number={9},
  pages={321},
  year={1964},
  publisher={APS}
}

@article{bornmann2020disruption,
author = {Bornmann, Lutz and Devarakonda, Sitaram and Tekles, Alexander and Chacko, George},
    title = {Are disruption index indicators convergently valid? The comparison of several indicator variants with assessments by peers},
    journal = {Quant. Sci. Stud.},
    volume = {1},
    number = {3},
    pages = {1242-1259},
    year = {2020},
    month = {08},
    issn = {2641-3337},
    doi = {10.1162/qss_a_00068},
    url = {https://doi.org/10.1162/qss\_a\_00068},
  
}

@article{wu2019large,
 title = {Large teams develop and small teams disrupt science and technology},
	volume = {566},
	issn = {1476-4687},
	url = {https://doi.org/10.1038/s41586-019-0941-9},
	doi = {10.1038/s41586-019-0941-9},
	number = {7744},
	journal = {Nature},
	author = {Wu, Lingfei and Wang, Dashun and Evans, James A.},
	month = feb,
	year = {2019},
	pages = {378--382},
}

@article{chu2021slowed,
  author = {Johan S. G. Chu  and James A. Evans },
title = {Slowed canonical progress in large fields of science},
journal = {Proc. Natl Acad. Sci. USA},
volume = {118},
number = {41},
pages = {e2021636118},
year = {2021},
doi = {10.1073/pnas.2021636118},
URL = {https://www.pnas.org/doi/abs/10.1073/pnas.2021636118}}

@book{arthur2009nature,
  title={The nature of technology: What it is and how it evolves},
  author={Arthur, W Brian},
  year={2009},
  publisher={Simon and Schuster}
}

@article{uzzi2013atypical,
author = {Brian Uzzi  and Satyam Mukherjee  and Michael Stringer  and Ben Jones },
title = {Atypical Combinations and Scientific Impact},
journal = {Science},
volume = {342},
number = {6157},
pages = {468-472},
year = {2013},
doi = {10.1126/science.1240474},
URL = {https://www.science.org/doi/abs/10.1126/science.1240474}}
\end{refsection}
\end{document}